\documentclass[twocolumn,showpacs,amsmath,amssymb,pra,superscriptaddress]{revtex4-1}
\usepackage{graphicx}
\usepackage{dcolumn}
\usepackage{bm}
\usepackage {longtable}
\usepackage{color}      
\newcommand{\ket}[1]{|#1\rangle}

\begin{document}

\title{Jaynes-Cummings dynamics in mesoscopic ensembles of  Rydberg-blockaded atoms}

\author{I.~I.~Beterov}
\email{beterov@isp.nsc.ru}
\affiliation{Rzhanov Institute of Semiconductor Physics SB RAS, 630090 Novosibirsk, Russia}
\affiliation{Novosibirsk State University, 630090 Novosibirsk, Russia}

\author{T.~Andrijauskas}
\affiliation{Institute of Theoretical Physics and Astronomy, Vilnius University, Gostauto 12, LT-01108 Vilnius, Lithuania}

\author{D.~B.~Tretyakov}
\affiliation{Rzhanov Institute of Semiconductor Physics SB RAS, 630090 Novosibirsk, Russia}

\author{V.~M.~Entin}
\affiliation{Rzhanov Institute of Semiconductor Physics SB RAS, 630090 Novosibirsk, Russia}

\author{E.~A.~Yakshina}
\affiliation{Rzhanov Institute of Semiconductor Physics SB RAS, 630090 Novosibirsk, Russia}
\affiliation{Novosibirsk State University, 630090 Novosibirsk, Russia}
\affiliation{Russian Quantum Center, Skolkovo, Moscow Region, 143025, Russia}

\author{I.~I.~Ryabtsev}
\affiliation{Rzhanov Institute of Semiconductor Physics SB RAS, 630090 Novosibirsk, Russia}
\affiliation{Novosibirsk State University, 630090 Novosibirsk, Russia}
\affiliation{Russian Quantum Center, Skolkovo, Moscow Region, 143025, Russia}

\author{S.~Bergamini}
\affiliation{The Open University, Walton Hall, MK7 6AA, Milton Keynes, United Kingdom}

\date{31 October 2014}

\begin{abstract}

We show that Jaynes-Cummings dynamics can be observed in mesoscopic atomic ensembles interacting with a classical
electromagnetic field in the regime of a Rydberg blockade where the time dynamics of the average number of Rydberg
excitations in  mesoscopic ensembles displays collapses and revivals typical of this model.  As the frequency of Rabi
oscillations between collective states of Rydberg-blockaded ensembles depends on the number of
interacting atoms, for randomly loaded optical dipole traps we predict collapses and revivals of Rabi oscillations.
We have studied the effects of finite interaction strengths and a finite laser line width on the visibility of the
revivals. We have shown that observation of collapses and revivals of Rabi oscillations can be used as a signature of the Rydberg blockade without the need to measure the exact number of Rydberg atoms. 

\end{abstract}

\maketitle

\section{Introduction}
\begin{figure}[!t]
\includegraphics[width=\columnwidth]{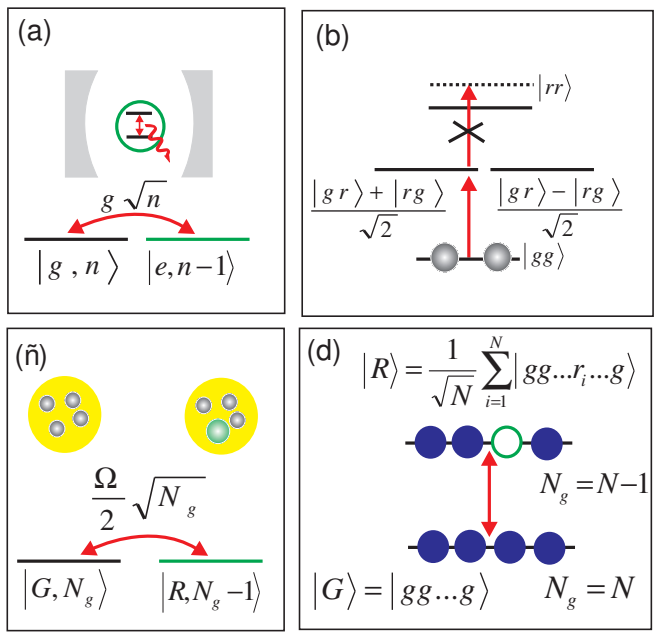}
\vspace{-.5cm}
\caption{
\label{Scheme}(Color online)
(a) Interaction of a single atom with the quantized electromagnetic field in a cavity is described by the Jaynes-Cummings model~\cite{Jaynes1963}. Two coupled atom-field states are $\ket{g,n}$ and $\ket{e,n-1}$ where $n$ is the number of photons and $\ket{g},\ket{e}$ are ground and excited states of the atom; (b) Rydberg blockade for two interacting atoms. The shift in the collective energy level $\ket{rr}$ caused by the interaction prohibits excitation of two Rydberg atoms by a narrow-band laser radiation; (c) scheme of the coupled states of a mesoscopic atomic ensemble with the number of ground-state atoms $N_g$ is considered as being equivalent to the number of photons in the Jaynes-Cummings model; (d) scheme of the collective states of the atomic ensemble
with $N$ atoms interacting with laser radiation in the regime of the Rydberg blockade.
}
\end{figure}

The Jaynes-Cummings model~\cite{Jaynes1963} is a basic model of interaction of two-level quantum systems with a
single-mode quantized electromagnetic field. It has been used to describe a variety of quantum systems including
neutral atoms in a cavity~\cite{Rempe1987} as shown in Fig.~\ref{Scheme}(a), single trapped ions~\cite{Leibfried2003},
quantum dots~\cite{Meier2004, Basset2013} and graphene~\cite{Dora2009}.

Quantum fluctuations in the number of photons in a mode of the quantized electromagnetic field lead to nonclassical features
in time dynamics of a two-level system  known as collapses and revivals of Rabi oscillations \cite{Narozhny1981}.
The frequency of Rabi oscillations between states of a two-level quantum system is proportional to $\sqrt{n}$, where
$n$ is the number of photons in the mode of the electromagnetic field which is resonant to the atomic transition. For a
coherent state of the electromagnetic field which has a randomly distributed number of photons, dephasing and
consequent rephasing of Rabi oscillations with different frequencies are observed in the Jaynes-Cummings model. This dynamics has been theoretically
predicted~\cite{Narozhny1981} and experimentally demonstrated for the one-atom maser~\cite{Rempe1987} and for a single
trapped ion~\cite{Leibfried2003}.

In this paper we show that similar dynamics of Rabi oscillations could be observed in mesoscopic atomic
ensembles interacting with  resonant laser radiation in the regime of the Rydberg blockade. Strong interaction between
Rydberg atoms in an optical dipole trap leads to the Rydberg blockade phenomenon~\cite{Lukin2001, Comparat2010}
when only one atom in the ensemble within the blockade radius can be excited to a Rydberg state due to the shift in collective energy levels as
illustrated in Fig.~\ref{Scheme}(b) for two atoms. In the regime of the perfect Rydberg blockade the mesoscopic atomic
ensemble is effectively a two-level system with two levels represented by collective Dicke states $\ket{G}$ and
$\ket{R}$~\cite{Lukin2001}, shown in  Fig.~\ref{Scheme}(d),
\begin{eqnarray}
\label{eq1}
\ket {G}  &=&\ket {g...g} \nonumber \\
\ket {R}  &=& \frac{{1}}{{\sqrt {N}} }\sum\limits_{i = 1}^{N}{\ket {g...r_{i} ..g} }.
\end{eqnarray}

\noindent Here $N$ is the number of atoms, $\ket{g}$  is the ground, and $\ket{r}$ is a Rydberg state of a single
atom. In the excited state $\ket{R}$ one Rydberg excitation is shared between all atoms in the ensemble. In  such
blockaded  ensembles, also called "superatoms"~\cite{Stanojevic2009, Lukin2001, Garttner2012}, Rabi oscillations between
collective states have been experimentally observed~\cite{Dudin2012}. Enhancement of the
atom-light interaction strength in Rydberg blockaded ensembles is of particular interest for cavity quantum
electrodynamics~\cite{Guerlin2010} and nonlinear optics with single photons~\cite{Firstenberg2013,Firstenberg2013a}. This enhancement results in the increased  frequency of Rabi oscillations of single Rydberg excitation by a
factor of $\sqrt N$ compared to the single-atom Rabi frequency. This is equivalent to
the dependence of the Rabi frequency in the Jaynes-Cummings model on the number of photons. In this paper we show that
collapses and revivals of Rabi oscillations in atomic ensembles randomly loaded in optical dipole traps~\cite{Grimm2000}
and optical lattices can be observed for realistic experimental parameters with dynamics that follow the
Jaynes-Cummings model.

Fluctuations in the number of trapped atoms have been known as an obstacle for encoding of quantum information in the collective states of the mesoscopic atomic ensembles and implementation of quantum logic gates~\cite{Lukin2001, Saffman2010}. The dephasing of Rabi oscillations can be partially suppressed by the use of ensembles containing several hundreds of atoms~\cite{Dudin2012}, which corresponds to the classical limit in our model. Recently we have proposed the schemes of deterministic single-atom Rydberg excitation and quantum gates with mesoscopic ensembles containing random numbers of strongly interacting atoms in the regime of the Rydberg blockade~\cite{Beterov2011, Beterov2013, Beterov2014}. The mesoscopic atomic ensembles containing random numbers of atoms can be considered as single qubits, and the quantum gates can be implemented using adiabatic laser excitation and the Rydberg blockade.

The observation of coherent many-body Rabi oscillations is a prerequisite for implementation of quantum gates based on the Rydberg blockade~\cite{Isenhower2010, Wilk2010,Urban2009}. However, due to fluctuations in the collective Rabi frequency, these oscillations cannot be observed in mesoscopic atomic ensembles with a random number of atoms. Although the presence of a single atom in the trap can be detected using resonant fluorescence~\cite{Isenhower2010, Wilk2010, Nogrette2014}, it is difficult to measure the number of atoms in the mesoscopic ensemble loaded into a single optical dipole trap~\cite{Hume2013}. This can be an obstacle to verify that the experimental conditions are suitable for quantum gates and there is no additional dephasing caused by the finite linewidth of laser radiation or stray electric or magnetic fields, and the interaction between atoms is strong enough to reach the regime of the perfect Rydberg blockade. 

Another common difficulty for observation of the Rydberg blockade is the measurement of the number of Rydberg atoms. The detection of single Rydberg excitations in the experiment can be implemented using a variety of methods. In the experiments with single trapped atoms the losses of Rydberg atoms from the optical dipole trap can be used as a signature of successful Rydberg excitation~\cite{Urban2009}. In the atomic ensembles single Rydberg excitations can be observed using a laser pulse which drives a Rydberg atom to a low-excited state with a short radiative lifetime. The spontaneously emitted photons can be detected using single-photon detectors with sufficiently high quantum efficiency~\cite{Dudin2012}. Rydberg atoms can also be  detected in a more straightforward  way using selective field ionization in a dc electric field and microchannel plates or channeltron electron multipliers~\cite{Stebbings1975,Ryabtsev2007}. In the latter case it is possible to measure the actual number of the detected Rydberg atoms~\cite{Ryabtsev2010}. However, accurate determination of the number of Rydberg atoms in atomic ensembles remains a challenging problem. For low detection efficiencies it is difficult to distinguish events when one or two Rydberg atoms are excited~\cite{Ryabtsev2007, Ryabtsev2010}, which is of crucial importance for observation of Rydberg blockade~\cite{Lukin2001} and application of Rydberg atoms to quantum information~\cite{Saffman2010}. In this paper we show that collapses and revivals of Rabi oscillations in the atomic ensembles due to fluctuations of the number of trapped atoms can be used as a signature of the Rydberg blockade without the need to measure the number of Rydberg excitations within the ensemble. 

\section{Jaynes-Cummings model and mesoscopic atomic ensembles}

The Jaynes-Cummings Hamiltonian is written as~\cite{Jaynes1963}

\begin{equation}
\label{eq2}
\hat {H}_{JC} = \frac{{1}}{{2}}\hbar \omega _{0} \hat {\sigma}_{z} + \hbar
\omega \hat {a}^{\dagger} \hat {a} + \hbar g\left( {\hat {a}^{\dagger} \hat {\sigma
}^{-}  + \hat {\sigma} ^{ +} \hat {a}} \right),
\end{equation}

\noindent where $\omega _{0} $ is the transition frequency between two atomic states, $\omega $ is the frequency of
laser radiation, $g$ is the coupling strength, $\hat{\sigma}^{-} $ and $\hat{\sigma}^{+} $ are the lowering and
raising operators for the two-level atom, whereas $\hat{a}$ and $\hat{a}^{\dagger} $ are annihilation and creation operators
for the quantized electromagnetic field. In the following we assume that laser radiation is resonant with atomic
transition, i.e., $\omega_0 =\omega$.

For a two-level atom in a quantized field,  collective atom-field state $\ket{g,n}$ corresponds to the case when the atom is in ground
state $\ket{g}$ and the mode of the field contains $n$ photons. This state  is coupled to state $\ket{e,n-1}$, which corresponds to the case when the atom is in excited state $\ket{e}$ and the field contains $n-1$ photons, as shown in Fig.~\ref{Scheme}(a). 
The Rabi frequency depends on the number of photons and on the coupling strength as $g \sqrt{n}$.

The atom is initially prepared in state $\ket{g}$, and the field is in state $\sum\limits_{n=0}^{\infty}{C\left( n
\right) \ket{n}}$, where the probability to find $n$ photons in the field mode is $p(n)=|C(n)|^2$. The number of photons in
the coherent state of the electromagnetic field is described by a Poissonian distribution~\cite{Loudon1983}, 
\begin{equation}
\label{eq3}
p\left(
{n} \right) = \left( {\bar {n}} \right)^{n}\mathrm{exp}\left( { - \bar {n}} \right)/\left( {n!} \right).
\end{equation}
\noindent Solving the
Schr\"odinger equation yields probability $P_e$ to find the atom in the excited state, as given
in~Refs.\cite{Brune1996,Gea-Banacloche1990},

\begin{equation}
\label{eq4}
P_e = \sum\limits_{n=1}^{\infty} {p\left( {n} \right)\mathrm{sin}^{2}\left( {g t\sqrt {n}}
\right)} .
\end{equation}

\noindent
The $\sqrt{n}$ dependence in
Eq.(\ref{eq4}) arises from matrix elements $\left\langle {n} \right|\hat{a}^{ +} \left|{n-1} \right\rangle =
\left\langle {n-1} \right|\hat{a}\left| {n} \right\rangle = \sqrt {n}$ of creation and annihilation operators.
Similarly, due to the dependence of the Rabi frequency of single-atom excitation in Rydberg-blockaded ensembles on the number of atoms,
the probability of single-atom excitation in an ensemble with a random number of atoms is described  by

\begin{equation}
\label{eq5}
P_1 = \sum\limits_{N} {p\left( {N} \right)\mathrm{sin}^{2}\left( {\sqrt{N}\Omega t/2}
\right)} .
\end{equation}

\noindent where $p(N)$ is the probability to have $N$ atoms in the ensemble. 
Given the analogy of Eqs.~(\ref{eq4}) and (\ref{eq5}), we can introduce a new quantum number for the mesoscospic ensemble of atoms
which is equivalent to the number of photons $n$ and is described by similar annihilation and creation operators, as
follows.

States of the atomic ensemble with $N$ atoms can be described as $\ket{G,N_g}$ and $\ket{R,N_g}$, where $N_g$ is the
number of ground-state atoms in the ensemble as shown in Fig.~\ref{Scheme}(d).  The excitation of the ensemble to a state
with a single Rydberg excitation can be treated as reducing of $N_g$ by one, which is equivalent to a single-photon
absorption in the Jaynes-Cummings model. States $\ket{G,N_g}$ and $\ket{R,N_g-1}$ are degenerate in energy and coupled, as shown in
Fig.~\ref{Scheme}(c). The $\sqrt{N_g}=\sqrt{N}$ dependence of the Rabi frequency in the atomic ensemble arises because
of the matrix element of the electric dipole transition of the ensemble to the Rydberg state. This is $\hat {V}_{las}(t) =
- \sum\limits_{i = 1}^{N} {\hat {\mu} ^{\left( {i} \right)}E\rm{cos\left(\omega t\right)}} $, where $\hat{\mu} ^{\left( {i} \right)}$ is the dipole
transition operator for the $i$th atom in the ensemble and $E$ is the amplitude of the electric field. We can write matrix
elements of equivalent creation $\hat{a}^{\dagger}$ and annihilation $\hat{a}$ operators for the states of a superatom,

\begin{eqnarray}
\label{eq6}
\hbar \Omega \left\langle {N_g} \right|\hat {a}^{\dagger}  \left| {N_g - 1}\right\rangle &=& - \frac{{1}}{{\sqrt {N}} }\sum\limits_{i = 1}^{N} {\mu_{gr}^{\left( {i} \right)}E} = -\sqrt{N}\mu_{gr}E \nonumber \\
\hbar\Omega\left\langle{N_g} \right|\hat {a}^{\dagger}\left| {N_g-1}\right\rangle&=& \hbar\Omega\left\langle{N_g-1}\right|\hat{a}\left|{N_g}\right\rangle,
\end{eqnarray}

\noindent where $\mu_{gr}^{\left( {i} \right)} = \mu_{gr}$ is the matrix element of the dipole moment for the transition
between ground and Rydberg states.

The effective Hamiltonian of this problem in the rotating-wave approximation is written in the following form:

\begin{equation}
\label{eq7}
\hat {H}_{JCA} =  \left(\hbar\Omega/2\right)
 \left( {\hat {a}^{\dagger}  \hat {\sigma}^{-} + \hat {a}\hat {\sigma}^{+}}  \right).
\end{equation}

\noindent Here $\hat{a}$ and $\hat{a}^{\dagger} $ are annihilation and
creation operators, and $\sigma^{+}=\ket{R}\langle G |$ and $\sigma^{-}=\ket{G}\langle R |$ are the raising and lowering operators for collective states $\ket{G}$ and $\ket{R}$.

In quantum optics, a coherent state of the electromagnetic field in the basis of Fock states is a superposition,
\begin{equation}
\label{eq8}
\ket{\alpha}=\sum\limits_{n=0}^{\infty}{\mathrm{exp}\left(-\left|\alpha\right|^2/2\right)\frac{\alpha^n}{\sqrt{n!}}\ket{n}}.
\end{equation}
\noindent Here $\left|\alpha\right|^2=\bar{n}$, where $\bar{n}$ is the average number of photons. However, due to phase fluctuations and technical noise, lasers often produce statistical mixtures of number states rather than pure coherent states~\cite{Louisell1973,Cohen-Tannoudji2004}. These statistical mixtures are described by the density matrix which has only diagonal terms,

\begin{equation}
\label{eq9}
\hat{\rho}=\sum\limits_{n=0}^{\infty}{p(n)\ket{n}\left\langle n\right|}.
\end{equation}

\noindent
When atoms interact with either type of states, described by Eqs.~(\ref{eq8}) and (\ref{eq9}), similar dynamics of collapses and revivals is observed. However, atomic ensembles considered in this paper cannot be described by a quantum superposition of states with different numbers of atoms in analogy with Eq.~(\ref{eq8}). Therefore, our simulation of Jaynes-Cummings dynamics in mesoscopic ensembles corresponds to the statistical mixture given by Eq.~(\ref{eq9}) and must be considered as being semiclassical. This is in agreement with the conclusion of Ref.~\cite{Molmer1997} that observable effects in quantum optics do not depend on the existence of quantum optical coherences.  

The quantum properties of laser radiation are not considered in the present study. The above
consideration is valid only for two isolated levels of a quantum
oscillator which corresponds to Jaynes-Cummings dynamics in the
rotating-wave approximation.

\section{Numeric simulation}

We have numerically calculated the probabilities $q_{Ry}(n)$  to excite $n$ Rydberg atoms in the mesoscopic atomic ensemble by solving the equations for the probability amplitudes, derived from the Schr\"odinger equation. We considered a randomly loaded optical dipole trap with an average number of atoms $\bar{N}=7$ which is close to the number of atoms in the mesoscopic ensembles, considered in our previous papers~\cite{Beterov2011, Beterov2013, Beterov2014}. We assumed that the trapping light is switched off prior to Rydberg excitation and single-atom Rabi frequency at $\ket{g} \to \ket{r}$ transition is $\Omega /\left( {2\pi}  \right) = 1$~MHz.

The many-body Hamiltonian for a mesoscopic ensemble interacting with laser radiation and with binary atom-atom interactions taken into account is written as~\cite{Stanojevic2009}.
\begin{equation}
\label{eq10}
\hat {H}_{M} = \frac{{1}}{{2}}\hbar \Omega \sum\limits_{i = 1}^{N} {\left(
{\hat {\sigma} _{gr}^{\left( {i} \right)} + \hat {\sigma} _{rg}^{\left( {i}
\right)}}  \right)} + \sum\limits_{i = 1,i < j}^{N} {V_{ij}}  \hat {\sigma
}_{rr}^{\left( {i} \right)} \hat {\sigma} _{rr}^{\left( {j} \right)} .
\end{equation}

\noindent Here $V_{ij} = {{\hbar C_{6}}  \mathord{\left/ {\vphantom {{\hbar C_{6}}
{R_{ij}^{6}} }} \right. \kern-\nulldelimiterspace} {R_{ij}^{6}} }$ is the interaction strength for the van der Waals interaction, and
$R_{ij} $ is the spatial separation of a pair of atoms $i$ and $j$. The $\hat {\sigma} _{ab}^{\left( {i}
\right)} = \left| {a_{i}}  \right\rangle \left\langle {b_{i}}  \right|$
operators refer to the transition between states $\ket{a}$ and
$\ket{b}$ of the $i$th atom. These states can be either a ground $\ket{g}$ or a Rydberg $\ket{r}$
state. The atoms are randomly located inside the optical dipole trap;
coordinates $x_{i} ,y_{i} ,z_{i} $ of each atom are described by the
normal distribution with standard deviation $r$. The interaction
strength for the Cs 80$S$ state used in the simulations is ${{C_6}
\mathord{\left/ {\vphantom {{C_{6}}  {\left( {2\pi}  \right)}}} \right.
\kern-\nulldelimiterspace} {\left( {2\pi}  \right)}} = 3.2 \times
10^{6}\,\rm{MHz}\,\mu \rm m^{6}$ (see Supplemental Material for Ref.~\cite{Beterov2013}). The blockade radius is obtained by equating the energy of the van der Waals interaction with the collective Rabi frequency~\cite{Honer2011}: $R = \left(
{{{C_{6}}  \mathord{\left/ {\vphantom {{C_{6}}  {\Omega} }} \right.
\kern-\nulldelimiterspace} {\Omega} }\sqrt {N}}  \right)^{1/6} \approx 10\,\mu \rm m$ for
$N=7$~atoms. 

\begin{figure}[!t]
\includegraphics[width=\columnwidth]{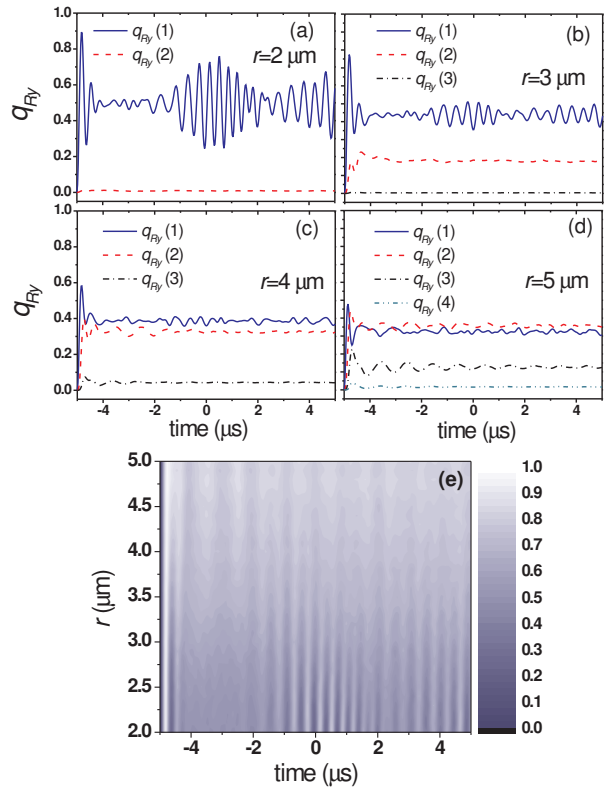}
\vspace{-.5cm}
\caption{
\label{qRy}(Color online)
The numerically calculated dependencies of the probabilities $q_{Ry}(n)$  to excite $n$ Rydberg atoms in a mesoscopic
ensemble with $\bar{N} = 7$ atoms, randomly distributed in the optical dipole trap with the radius (a) $r=2\,\mu \rm m$, 
(b) $r=3\,\mu \rm m$, (c) $r=4\,\mu \rm m$,  and (d) $r=5\,\mu \rm m$. (e) The calculated density plot of the probability to excite at least one Rydberg atom $P_{Ry}$  as a function of $r$ and time.
}
\end{figure}

The dynamics of excitation to a Rydberg state in the ensemble is shown in Fig.~\ref{qRy}. These results are averaged over random spatial configurations of atoms within the trap. The equations for the probability amplitudes, derived from Schr\"odinger equation, are first solved for  collective states in mesoscopic ensembles containing up to $N_{max}$=20 atoms for 20,000 different
spatial configurations  drawn randomly and are then averaged using the Poissonian
distribution of the number of atoms with  $\bar {N} = 7$ atoms. A similar approach has been used in our previous papers~\cite{Ryabtsev2010a, Beterov2011, Beterov2013}. The probability to have more than $N_{max}=20$~atoms for the Poissonian
distribution with  $\bar {N} = 7$  does not exceed $3.1 \times 10^{-4}$.
Following Ref.~\cite{Stanojevic2009}, we have limited the
number of possible Rydberg excitations to 2 for $r = 2\,\mu \rm m$ and to 3 for $r =
3\,\mu \rm m$. We have checked out that results of our simulations for small values of $r$ are not affected
by this approximation.

The collapses and revivals are clearly observed in the simulated dynamics of the probability of single-atom Rydberg excitation $q_{Ry}(1)$ for $r = 2\,\mu \rm m$ on a timescale of $10 \,\mu \rm s$, which is shown as a solid curve in Fig.~\ref{qRy}(a).
The probability $q_{Ry}(2)$ to excite two Rydberg atoms [dashed curve in Fig.~\ref{qRy}(a)] is close to zero, showing that Rydberg blockade is perfect. The calculated dynamics of $q_{Ry}(1)$ is equivalent to the genuine Jaynes-Cummings dynamics, described by Eqs.~\ref{eq3} and \ref{eq4}. For $r = 3\,\mu \rm m$, despite the fact that the size of the trap is still smaller than 
the blockade radius, the revivals in $q_{Ry}(1)$  [solid curve in Fig.~\ref{qRy}(b)] are suppressed due to a partial breakdown of Rydberg blockade caused by a reduced average energy of the van der Waals interaction of atoms in the trap. The breakdown of Rydberg blockade is observed as an increase in the probability to excite two Rydberg atoms $q_{Ry}(2)$ [dashed curve in Fig.~\ref{qRy}(b)]. For larger sizes of the optical dipole traps, shown in Figs.~\ref{qRy}(c) and \ref{qRy}(d), the revivals are not observed. Oscillations in the time dependencies of $q_{Ry}(3)$ and $q_{Ry}(4)$ in Fig.~\ref{qRy}(d) result from coherent interaction of uncorrelated atoms with laser radiation. The large number of excited atoms corresponds to a weak van der Waals interaction within the ensemble. For these atoms independent Rabi oscillations at a single-atom Rabi frequency become observable.

From the calculated probabilities $q_{Ry}(n)$ it is easy to find the average number of Rydberg excitations $N_{Ry}$ and the probability to have at least one Rydberg excitation $P_{Ry}$,
\begin{eqnarray}
\label{eq11}
 N_{Ry}=\sum\limits_{n=1}^{N_{max}}{n q_{Ry}(n)} \\\nonumber
 P_{Ry}=\sum\limits_{n=1}^{N_{max}}{q_{Ry}(n)}.
\end{eqnarray}
\noindent
In the case of the perfect blockade, shown in Fig.~\ref{qRy}(a), $N_{Ry}$ is equal both to the probability to have exactly one Rydberg excitation $P_1$ and to the probability to have at least one Rydberg excitation $P_{Ry}$. The latter refers to the low-efficiency detector in the counting mode when the experimental conditions where single-atom and two-atom  (and more) events cannot be distinguished~\cite{Dudin2012}.

\begin{figure}[!t]
\includegraphics[width=\columnwidth]{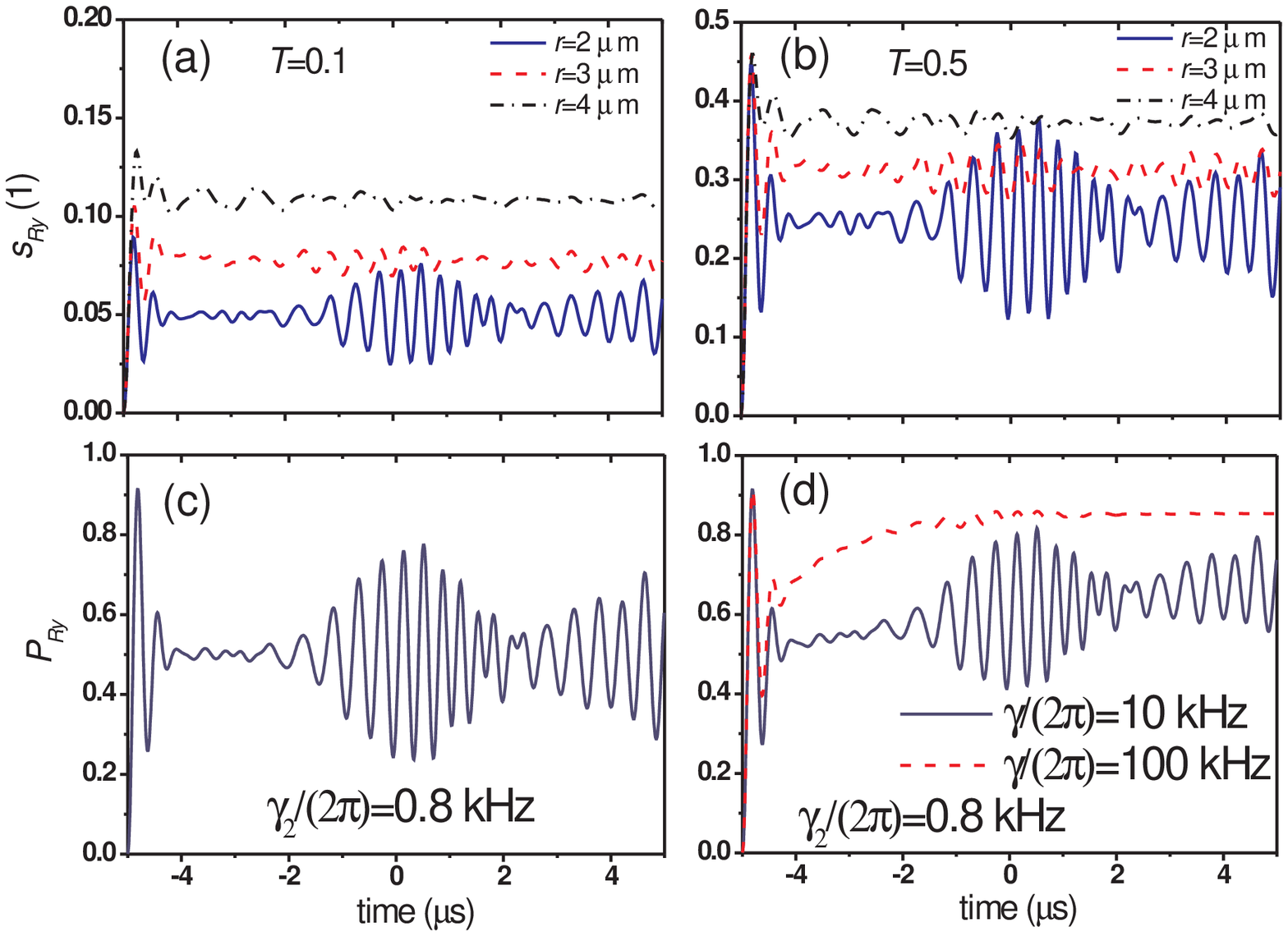}
\vspace{-.5cm}
\caption{
\label{Density}
(Color online) The numerically calculated dependencies of the probability $s_{Ry}(1)$ to detect a single Rydberg atom 
in a mesoscopic ensemble with $\bar{N} = 7$ atoms for finite detection efficiency (a) $T=0.1$ and (b) $T=0.5$, using Eq.~(\ref{eq12}); the numerically calculated probabilities to excite at least one Rydberg atom $P_{Ry}$ calculated using Eqs.~\ref{eq13}-\ref{eq17} taking into account the finite line width of the Rydberg Cs $80S$ state $\gamma_2/(2\pi)=0.8$~kHz and laser line width 
(c) $\gamma/(2\pi)=0$~kHz; (d) $\gamma/(2\pi)=10$~kHz (solid curve) and $\gamma/(2\pi)=100$~kHz (dashed curve).
}
\end{figure}

Figure~\ref{qRy}(e) shows the density plot of $P_{Ry}$, calculated using~Eq.~(\ref{eq10}) as a function of time and radius of the optical dipole trap $r$. A breakdown of Rydberg blockade for $r>2\, \mu \rm m$ leads to strong suppression of the revivals. This implies that Jaynes-Cummings dynamics in randomly loaded mesoscopic atomic ensembles can be used as a signature of perfect Rydberg blockade without the need to measure the exact number of Rydberg excitations.

We have studied the effect of the finite detection efficiency $T$ on the observation of the collapses and revivals in the probability of single-atom excitation. The probability $s_{Ry}(k)$ to detect $k$ atoms is related to the probabilities $q_{Ry}(i)$ to excite $i$ atoms as~\cite{Ryabtsev2007}:

\begin{equation}
\label{eq12}
s_{Ry}\left( {k} \right) = \sum\limits_{i = k}^{\infty}  {T^{k}\left( {1 - T} 
\right)^{i - k}C_{i}^{k} q_{Ry} \left( {i} \right).} 
\end{equation}

\noindent The results of a numeric calculation of $s_{Ry}(1)$ are shown in Figs.~\ref{Density}(a) and \ref{Density}(b) for detection efficiencies $T=0.1$ and $T=0.5$, respectively, using the excitation probabilities $q_{Ry}(i)$ from Fig.~\ref{qRy}. The collapses and revivals are clearly visible at $r=2 \,\mu \rm m$ (solid curve) and almost disappear with a blockade breakdown at $r=4 \,\mu \rm m$ (dashed-dotted curve) which shows that even for small detection efficiency $T=0.1$ it is possible to verify whether the regime of perfect blockade is achieved. 

We have also investigated the effect of the finite line width of the atomic transition on
collapses and revivals of oscillations in the dynamics of excitation. To simulate its effect on a time dependence of Rydberg excitation, but assuming a perfect blockade, we used the superatom
model~\cite{Stanojevic2009} which describes the whole ensemble in the regime of the perfect blockade as a two-level system
taking into account spontaneous decay of the excited state interacting with laser radiation, in our case  for the
cesium $80S$ state at the ambient temperature of 300~K this decay rate is $\gamma_2 /\left( {2\pi}  \right) =
0.8$~kHz~\cite{Beterov2009}.

We have solved the master equation for the mesoscopic atomic ensemble consisting of $N$ atoms interacting with the laser radiation in the regime of the Rydberg blockade~\cite{Petrosyan2013}, 

\begin{equation}
\label{eq13}
\dot {\hat {\rho} }\left( {t} \right) = - \frac{{i}}{{\hbar} }\left[ {\hat {H_B},\hat {\rho} 
\left({t} \right)} \right] + \hat {L}\hat {\rho} \left( {t} \right).
\end{equation}

\noindent The Hamiltonian is written as

\begin{equation}
\label{eq14}
\hat {H}_B = \frac{{1}}{{2}}\hbar \Omega \sum\limits_{i = 1}^{N} {\left(
{\hat {\sigma} _{gr}^{\left( {i} \right)} + \hat {\sigma} _{rg}^{\left( {i}
\right)}}  \right)}.
\end{equation}

\noindent The regime of the Rydberg blockade is simulated by removing all multiple excitations from the system of equations for the density matrix similar to our previous paper~\cite{Beterov2013}. The Liouvillian acting on the density matrix $\hat{\rho}$ written for the collective states of the mesoscopic ensemble is expressed as

\begin{equation}
\label{eq15}
\hat {L}\hat {\rho}  = \sum\limits_{i=1}^N {\left( {\hat {L}_{\gamma} ^{\left( 
{i} \right)} \hat {\rho}  + \hat {L}_{eg}^{\left( {i} \right)} \hat {\rho} } 
\right).} 
\end{equation}

\noindent Here we take into account the finite lifetime $\tau$ of the Rydberg state by using $\gamma_2=1/\tau$,

\begin{equation}
\label{eq16}
\hat {L}_{eg}^{\left( {i} \right)} \hat {\rho}  = \frac{{\gamma _{2} 
}}{{2}}\left( {2\hat {\sigma} _{ge}^{\left( {i} \right)} \hat {\rho} \hat 
{\sigma} _{eg}^{\left( {i} \right)} - \hat {\rho} \hat {\sigma 
}_{ee}^{\left( {i} \right)} - \hat {\sigma} _{ee}^{\left( {i} \right)} \hat 
{\rho} } \right),
\end{equation}

\noindent  and purely off-diagonal decay $\gamma$ due to the finite 
line width  of the laser radiation or technical noises,

\begin{equation}
\label{eq17}
\hat {L}_{\gamma} ^{\left( {i} \right)} \hat {\rho}  = \gamma \left( {2\hat 
{\sigma} _{ee}^{\left( {i} \right)} \hat {\rho} \hat {\sigma} _{ee}^{\left( 
{i} \right)} - \hat {\rho} \hat {\sigma} _{ee}^{\left( {i} \right)} - \hat 
{\sigma} _{ee}^{\left( {i} \right)} \hat {\rho} } \right).
\end{equation}

\noindent  The time-dependent single-atom excitation probability has been averaged over the Poissonian distribution with  $\bar{N} = 7$ atoms.  The results of the calculations are shown in Fig.~\ref{Density}(c). For $\gamma_2 /\left( {2\pi}  \right) =
0.8$~kHz and $\gamma=0$ the amplitude of the revivals is almost unchanged compared to Fig.~\ref{qRy}(a). The increase in
the laser line width to $\gamma/\left( {2\pi} \right) = 100$~kHz leads, however, to the strong suppression of the
revivals, as shown in Fig.~\ref{Density}(d) (dashed curve). However, if the laser line width is $\gamma/\left( {2\pi} \right) = 10$~kHz [solid curve in Fig.~\ref{Density}(d)], the amplitude of the revivals is moderately reduced compared to Fig.~\ref{Density}(c).
Surprisingly, pure off-diagonal decay results in the increase in the single-atom excitation probability up to 80\% [dashed curve in Fig.~\ref{Density}(d)].

Observation of collapses and revivals of the Rabi oscillations in the mesoscopic atomic ensembles could be a prerequisite for implementation of quantum logic gates, which we have proposed in our previous papers~\cite{Beterov2013,Beterov2014}, because both blockade breakdown and dephasing of the Rabi oscillations caused by a finite laser linewidth and technical noises will have a detrimental effect on the fidelity of the quantum gates.
\begin{figure}[!t]
\includegraphics[width=\columnwidth]{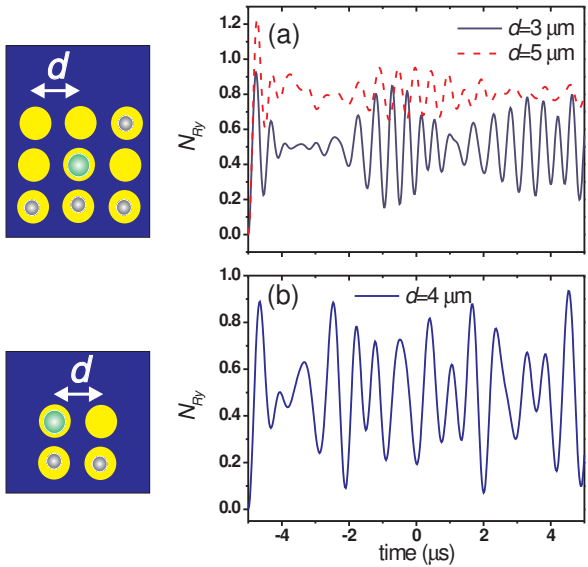}
\vspace{-.5cm}
\caption{
\label{Lattice}(Color online).
The time dependence of the average number of
Rydberg atoms excited in the optical lattice containing (a) nine and (b) four sites with 50\%
probability to load a single atom in each site.
}
\end{figure}

The Poissonian statistics of the atom-number distribution in the dipole trap is not always the necessary requirement for
observation of collapses and revivals. We have considered an array of
$N$ optical dipole traps loaded with single atoms in the regime of the
collisional blockade~\cite{Schlosser2002} with a $q=50\%$ probability of single-atom
occupancy for each trap~\cite{Reymond2003} and zero probability to load more than one atom. This is similar to the experimental conditions of the recent paper~\cite{Nogrette2014} where a spatial light modulator with the two-dimensional array of microtraps was used to create an array of optical dipole traps with arbitrary geometry. The minimal nearest-neighbor distance between the atoms in the array was $d=3\,\mu \rm m$.  The total
number $k$ of trapped atoms in the array is described by the binomial distribution:
$p\left( {k,N,q} \right) = C_N^k q^{k}\left( {1 - q} \right)^{N - k}$. We have simulated $N_{Ry}$ for $d=3\,\mu \rm m$  at $N$=9 and $q$=0.5 [solid curve in Fig.~\ref{Lattice}(a)] and $d=5\,\mu \rm m$ [dashed curve in Fig.~\ref{Lattice}(a)] . For the minimal nearest-neighbor distance $d=3\, \mu \rm m$ the collapses and revivals are clearly observed and the regime of the perfect Rydberg blockade is achieved. For smaller ensembles, considered in Ref.~\cite{Nogrette2014}, the Jaynes-Cummings dynamics is not observed, as shown in Fig.~\ref{Lattice}(b) for four atoms with $d=4\, \mu \rm m$  in the regime of perfect blockade.

\section{Interaction of two ensembles.} An interesting effect arises from the interplay in
the Jaynes-Cummings dynamics of two  interacting atomic ensembles. The spatially separated randomly loaded optical dipole traps are of interest for implementation of two-qubit quantum gates with mesoscopic atomic ensembles~\cite{Beterov2013, Beterov2014} or a deterministic quantum computation with one pure qubit (DQC1) algorithms~\cite{Mansell2014}. To implement the two-qubit gates based on mesoscopic atomic ensembles with a random number of atoms, discussed in our previous papers~\cite{Beterov2013, Beterov2014}, it is necessary to achieve the regime of a perfect blockade both within each ensemble and between two neighboring ensembles. Controlled rotations of the ensemble states, required for the DQC1 algorithm, must be performed in similar conditions, when laser excitation of Rydberg states within the ensemble could be blocked by Rydberg excitation of the separate control qubit~\cite{Mansell2014}.

The effective Hamiltonian for $N_{s}$ interacting superatoms is written as~\cite{Stanojevic2009}

\begin{equation}
\label{eq18}
\hat {H}_{S} = \frac{{1}}{{2}}\hbar \Omega \sum\limits_{j = 1}^{N_{s}}
{\sqrt {N_{j}}  \left( {\hat {\sigma} _{GR}^{\left( {j} \right)} + \hat
{\sigma} _{RG}^{\left( {j} \right)}}  \right)} + \sum\limits_{i = 1,i <
j}^{N_{s}}  {K_{ij} \hat {\sigma} _{RR}^{\left( {i} \right)} \hat {\sigma
}_{RR}^{\left( {j} \right)}}.
\end{equation}

\noindent Here we take into account the enhancement  of collective Rabi frequency for each ensemble. We consider the
mean  interaction strength between two ensembles, resulting from averaging the interaction energies between all
pairs of  atoms which belong to different superatoms~\cite{Stanojevic2009},

\begin{equation}
\label{eq19}
K_{ij} = \frac{{1}}{{N_{i} N_{j}} }\sum\limits_{p \in S_{i}}{\sum\limits_{q \in S_{j}}{V_{pq}} }.
\end{equation}

\noindent Here $N_i$ and $N_j$ are the numbers of atoms in $i$th and $j$th ensemble and $V_{pq}$ is the interaction strength between the $p$th atom from $i$th ensemble and $q$th atom from $j$th ensemble. Below we assume that $r \ll d$. In this case the differences in the interaction energy $V_{pq}$ for different $p$ and $q$ can be neglected  and the interaction part of the effective Jaynes-Cummings Hamiltonian is written as
\begin{eqnarray}
\label{eq20}
\hat {H}_2 = \left(\hbar\Omega/2\right)\left( \hat{a}_1^+  \hat {\sigma}_1^-+ \hat {a}_1\hat{\sigma}_1^+ \right)+&&\\
+\left(\hbar\Omega/2\right)\left( \hat{a}_2^+  \hat {\sigma}_2^-+ \hat {a}_2\hat{\sigma}_2^+ \right)
&+&K_{12}\sigma_1^+\sigma_1^-\sigma_2^+\sigma_2^-\nonumber.
\end{eqnarray}

\noindent Here the indices 1 and 2 correspond to the operators acting on different superatoms 1 and 2. 
We have calculated the
average number of Rydberg atoms excited in two interacting ensembles with $\bar {N} = 10$ atoms in each ensemble by solving the 
equations for the probability amplitudes $c_{MN}$ with the Hamiltonian, given by Eq.~(\ref{eq20}),

\begin{eqnarray}
\label{eq21}
i\dot{c}_{GG}  &= & \left(\Omega \sqrt{N}/2\right)\left(c_{RG}+c_{GR}\right) \nonumber \\ 
i\dot{c}_{GR}  &= & \left(\Omega \sqrt{N}/2\right)\left(c_{GG}+c_{RR}\right)=i\dot{c}_{RG}   \\ 
i\dot{c}_{RR}  &= & \left(\Omega \sqrt{N}/2\right)\left(c_{RG}+c_{GR}\right) + K_{12}c_{RR}. \nonumber
\end{eqnarray}

\begin{figure}[!t]
\includegraphics[width=\columnwidth]{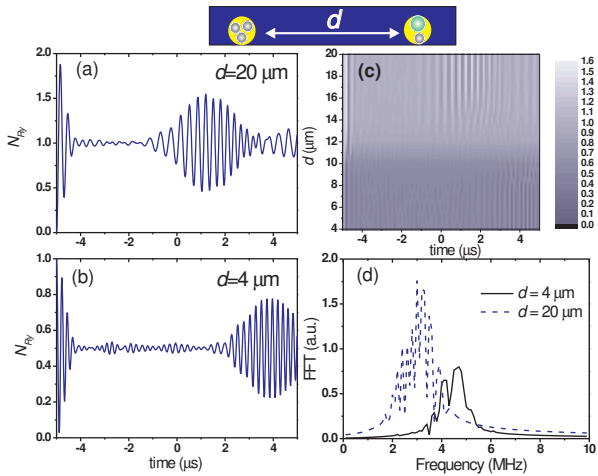}
\vspace{-.5cm}
\caption{
\label{Ensembles}
(Color online) Time dependence of the average number $N_{Ry}$ of
Rydberg atoms excited in two spatially separated, at distance $d$, mesoscopic atomic ensembles
with $\bar{N}=10$ atoms in each ensemble for (a)
$d=20\,\mu\rm m$ and (b) $d=4\, \mu \rm m$; (c) density plot of the average number of Rydberg atoms calculated
as a function of time and distance between ensembles; (d) the Fourier spectrum of the time dependencies of $N_{Ry}$ for
$d=4\,\mu\rm m$ (solid curve) and $d=20\, \mu \rm m$ (dashed curve).
}
\end{figure}

\noindent The ensembles are located in two optical dipole traps at distance $d$ between each
other. If the distance between traps is sufficiently large, the interaction becomes negligible as shown in
Fig.~\ref{Ensembles}(a) for $d = 20\,\mu \rm m$, and observed collapses and revivals correspond to the dynamics of a
single system with $\bar {N} = 10$ but with a doubled amplitude. If the distance between ensembles is small and the
regime of full blockade is reached within both ensembles, the dynamics of the system corresponds to a single superatom
with $\bar {N} = 20$ as shown in Fig.~\ref{Ensembles}(b) for $d = 4\,\mu \rm m$.

The dynamics of the Rydberg excitations in two mesoscopic ensembles located at arbitrary distances is illustrated in
Fig.~\ref{Ensembles}(c) as a density plot of $N_{Ry}$ calculated as a function of time and distance between optical
dipole traps and averaged over 500 samples with Poissonian distribution of the number of atoms in each trap. The
oscillations shown in Fig.~\ref{Ensembles}(c)(top part) correspond to the case of two non-interacting superatoms shown in
Fig.~\ref{Ensembles}(a),  whereas the full blockade between (and within) the ensembles is evident at the  bottom  of
Fig.~\ref{Ensembles}(c). For intermediate distances around $d = 9\,\mu \rm m$ the collapses and revivals disappear due to dephasing of the oscillations induced by the interaction between superatoms.  The Fourier spectrum of $N_{Ry}$ shown in Fig.~\ref{Ensembles}(d) illustrates the increase in the mean frequency of Rabi oscillations for $d = 4\,\mu \rm m$  due to the blockade of Rydberg excitation within two ensembles.

\section{Summary}
We have shown that strongly interacting mesoscopic atomic ensembles with random and unknown numbers of atoms, which are coupled to a classical
electromagnetic field, display the Jaynes-Cummings-type dynamics of single-atom laser excitation. The collapses and revivals of collective
oscillations between Dicke states of the atomic ensemble result from the $\sqrt{N}$ dependence of collective Rabi
frequency of single-atom excitation in the regime of the Rydberg blockade, where $N$ is the number of atoms. The interference of the Rabi oscillations with
different frequencies occurs due to the random loading of optical dipole traps or optical lattices. Due to $\sqrt{N}$ dependencies on the number of atoms these effects
are also relevant to the recent studies of superradiance~\cite{Akkermans2008,Garraway2011} and
subradiance~\cite{Bienaime2012} in atomic ensembles and to the investigation of Rydberg polaritons~\cite{Maxwell2013}.

An experimental observation of this effect can be used as a signature of the perfect Rydberg blockade without the need to measure the actual number of detected Rydberg atoms. This can be of great importance for quantum information with mesoscopic atomic ensembles containing a random number of atoms~\cite{Beterov2013, Beterov2014} where the Rydberg blockade within an atomic ensemble and between two ensembles is required for encoding of quantum information, implementation of two-qubit quantum gates, and DQC1 algorithms~\cite{Mansell2014}. Our approach could also be  useful for investigation of sub-Poissonian atom-number fluctuations in mesoscopic atomic ensembles~\cite{Whitlock2010}.

\begin{acknowledgments}
We thank A.~V.~Taichenachev, L.~V.~Il'ichev and T.~Lahaye, for helpful discussions.
This work was supported by the RFBR Grant No. 14-02-00680, by the Russian Academy of Sciences, by the EU FP7 IRSES Project "COLIMA", by the project TAP LLT 01/2012 of the Research Council of Lithuania, the National Science Council of Taiwan and by the Russian Quantum Center.
\end{acknowledgments}

%


\begin{thebibliography}{46}%
\makeatletter
\providecommand \@ifxundefined [1]{%
 \@ifx{#1\undefined}
}%
\providecommand \@ifnum [1]{%
 \ifnum #1\expandafter \@firstoftwo
 \else \expandafter \@secondoftwo
 \fi
}%
\providecommand \@ifx [1]{%
 \ifx #1\expandafter \@firstoftwo
 \else \expandafter \@secondoftwo
 \fi
}%
\providecommand \natexlab [1]{#1}%
\providecommand \enquote  [1]{``#1''}%
\providecommand \bibnamefont  [1]{#1}%
\providecommand \bibfnamefont [1]{#1}%
\providecommand \citenamefont [1]{#1}%
\providecommand \href@noop [0]{\@secondoftwo}%
\providecommand \href [0]{\begingroup \@sanitize@url \@href}%
\providecommand \@href[1]{\@@startlink{#1}\@@href}%
\providecommand \@@href[1]{\endgroup#1\@@endlink}%
\providecommand \@sanitize@url [0]{\catcode `\\12\catcode `\$12\catcode
  `\&12\catcode `\#12\catcode `\^12\catcode `\_12\catcode `\%12\relax}%
\providecommand \@@startlink[1]{}%
\providecommand \@@endlink[0]{}%
\providecommand \url  [0]{\begingroup\@sanitize@url \@url }%
\providecommand \@url [1]{\endgroup\@href {#1}{\urlprefix }}%
\providecommand \urlprefix  [0]{URL }%
\providecommand \Eprint [0]{\href }%
\providecommand \doibase [0]{http://dx.doi.org/}%
\providecommand \selectlanguage [0]{\@gobble}%
\providecommand \bibinfo  [0]{\@secondoftwo}%
\providecommand \bibfield  [0]{\@secondoftwo}%
\providecommand \translation [1]{[#1]}%
\providecommand \BibitemOpen [0]{}%
\providecommand \bibitemStop [0]{}%
\providecommand \bibitemNoStop [0]{.\EOS\space}%
\providecommand \EOS [0]{\spacefactor3000\relax}%
\providecommand \BibitemShut  [1]{\csname bibitem#1\endcsname}%
\let\auto@bib@innerbib\@empty
\bibitem [{\citenamefont {Jaynes}\ and\ \citenamefont
  {Cummings}(1963)}]{Jaynes1963}%
  \BibitemOpen
  \bibfield  {author} {\bibinfo {author} {\bibfnamefont {E.~T.}\ \bibnamefont
  {Jaynes}}\ and\ \bibinfo {author} {\bibfnamefont {F.~W.}\ \bibnamefont
  {Cummings}},\ }\href@noop {} {\bibfield  {journal} {\bibinfo  {journal}
  {Proceedings of the IEEE}\ }\textbf {\bibinfo {volume} {51}},\ \bibinfo
  {pages} {89} (\bibinfo {year} {1963})}\BibitemShut {NoStop}%
\bibitem [{\citenamefont {Rempe}\ \emph {et~al.}(1987)\citenamefont {Rempe},
  \citenamefont {Walther},\ and\ \citenamefont {Klein}}]{Rempe1987}%
  \BibitemOpen
  \bibfield  {author} {\bibinfo {author} {\bibfnamefont {G.}~\bibnamefont
  {Rempe}}, \bibinfo {author} {\bibfnamefont {H.}~\bibnamefont {Walther}}, \
  and\ \bibinfo {author} {\bibfnamefont {N.}~\bibnamefont {Klein}},\ }\href
  {\doibase 10.1103/PhysRevLett.58.353} {\bibfield  {journal} {\bibinfo
  {journal} {Phys. Rev. Lett.}\ }\textbf {\bibinfo {volume} {58}},\ \bibinfo
  {pages} {353} (\bibinfo {year} {1987})}\BibitemShut {NoStop}%
\bibitem [{\citenamefont {Leibfried}\ \emph {et~al.}(2003)\citenamefont
  {Leibfried}, \citenamefont {Blatt}, \citenamefont {Monroe},\ and\
  \citenamefont {Wineland}}]{Leibfried2003}%
  \BibitemOpen
  \bibfield  {author} {\bibinfo {author} {\bibfnamefont {D.}~\bibnamefont
  {Leibfried}}, \bibinfo {author} {\bibfnamefont {R.}~\bibnamefont {Blatt}},
  \bibinfo {author} {\bibfnamefont {C.}~\bibnamefont {Monroe}}, \ and\ \bibinfo
  {author} {\bibfnamefont {D.}~\bibnamefont {Wineland}},\ }\href {\doibase
  10.1103/RevModPhys.75.281} {\bibfield  {journal} {\bibinfo  {journal} {Rev.
  Mod. Phys.}\ }\textbf {\bibinfo {volume} {75}},\ \bibinfo {pages} {281}
  (\bibinfo {year} {2003})}\BibitemShut {NoStop}%
\bibitem [{\citenamefont {Meier}\ and\ \citenamefont
  {Awschalom}(2004)}]{Meier2004}%
  \BibitemOpen
  \bibfield  {author} {\bibinfo {author} {\bibfnamefont {F.}~\bibnamefont
  {Meier}}\ and\ \bibinfo {author} {\bibfnamefont {D.~D.}\ \bibnamefont
  {Awschalom}},\ }\href {\doibase 10.1103/PhysRevB.70.205329} {\bibfield
  {journal} {\bibinfo  {journal} {Phys. Rev. B}\ }\textbf {\bibinfo {volume}
  {70}},\ \bibinfo {pages} {205329} (\bibinfo {year} {2004})}\BibitemShut
  {NoStop}%
\bibitem [{\citenamefont {Basset}\ \emph {et~al.}(2013)\citenamefont {Basset},
  \citenamefont {Jarausch}, \citenamefont {Stockklauser}, \citenamefont {Frey},
  \citenamefont {Reichl}, \citenamefont {Wegscheider}, \citenamefont {Ihn},
  \citenamefont {Ensslin},\ and\ \citenamefont {Wallraff}}]{Basset2013}%
  \BibitemOpen
  \bibfield  {author} {\bibinfo {author} {\bibfnamefont {J.}~\bibnamefont
  {Basset}}, \bibinfo {author} {\bibfnamefont {D.-D.}\ \bibnamefont
  {Jarausch}}, \bibinfo {author} {\bibfnamefont {A.}~\bibnamefont
  {Stockklauser}}, \bibinfo {author} {\bibfnamefont {T.}~\bibnamefont {Frey}},
  \bibinfo {author} {\bibfnamefont {C.}~\bibnamefont {Reichl}}, \bibinfo
  {author} {\bibfnamefont {W.}~\bibnamefont {Wegscheider}}, \bibinfo {author}
  {\bibfnamefont {T.~M.}\ \bibnamefont {Ihn}}, \bibinfo {author} {\bibfnamefont
  {K.}~\bibnamefont {Ensslin}}, \ and\ \bibinfo {author} {\bibfnamefont
  {A.}~\bibnamefont {Wallraff}},\ }\href {\doibase 10.1103/PhysRevB.88.125312}
  {\bibfield  {journal} {\bibinfo  {journal} {Phys. Rev. B}\ }\textbf {\bibinfo
  {volume} {88}},\ \bibinfo {pages} {125312} (\bibinfo {year}
  {2013})}\BibitemShut {NoStop}%
\bibitem [{\citenamefont {D\'ora}\ \emph {et~al.}(2009)\citenamefont {D\'ora},
  \citenamefont {Ziegler}, \citenamefont {Thalmeier},\ and\ \citenamefont
  {Nakamura}}]{Dora2009}%
  \BibitemOpen
  \bibfield  {author} {\bibinfo {author} {\bibfnamefont {B.}~\bibnamefont
  {D\'ora}}, \bibinfo {author} {\bibfnamefont {K.}~\bibnamefont {Ziegler}},
  \bibinfo {author} {\bibfnamefont {P.}~\bibnamefont {Thalmeier}}, \ and\
  \bibinfo {author} {\bibfnamefont {M.}~\bibnamefont {Nakamura}},\ }\href
  {\doibase 10.1103/PhysRevLett.102.036803} {\bibfield  {journal} {\bibinfo
  {journal} {Phys. Rev. Lett}\ }\textbf {\bibinfo {volume} {102}},\ \bibinfo
  {pages} {036803} (\bibinfo {year} {2009})}\BibitemShut {NoStop}%
\bibitem [{\citenamefont {Narozhny}\ \emph {et~al.}(1981)\citenamefont
  {Narozhny}, \citenamefont {Sanchez-Mondragon},\ and\ \citenamefont
  {Eberly}}]{Narozhny1981}%
  \BibitemOpen
  \bibfield  {author} {\bibinfo {author} {\bibfnamefont {N.~B.}\ \bibnamefont
  {Narozhny}}, \bibinfo {author} {\bibfnamefont {J.~J.}\ \bibnamefont
  {Sanchez-Mondragon}}, \ and\ \bibinfo {author} {\bibfnamefont {J.~H.}\
  \bibnamefont {Eberly}},\ }\href {\doibase 10.1103/PhysRevA.23.236} {\bibfield
   {journal} {\bibinfo  {journal} {Phys. Rev. A}\ }\textbf {\bibinfo {volume}
  {23}},\ \bibinfo {pages} {236} (\bibinfo {year} {1981})}\BibitemShut
  {NoStop}%
\bibitem [{\citenamefont {Lukin}\ \emph {et~al.}(2001)\citenamefont {Lukin},
  \citenamefont {Fleischhauer}, \citenamefont {C\^ot\'e}, \citenamefont {Duan},
  \citenamefont {Jaksch}, \citenamefont {Cirac},\ and\ \citenamefont
  {Zoller}}]{Lukin2001}%
  \BibitemOpen
  \bibfield  {author} {\bibinfo {author} {\bibfnamefont {M.~D.}\ \bibnamefont
  {Lukin}}, \bibinfo {author} {\bibfnamefont {M.}~\bibnamefont {Fleischhauer}},
  \bibinfo {author} {\bibfnamefont {R.}~\bibnamefont {C\^ot\'e}}, \bibinfo
  {author} {\bibfnamefont {L.~M.}\ \bibnamefont {Duan}}, \bibinfo {author}
  {\bibfnamefont {D.}~\bibnamefont {Jaksch}}, \bibinfo {author} {\bibfnamefont
  {J.~I.}\ \bibnamefont {Cirac}}, \ and\ \bibinfo {author} {\bibfnamefont
  {P.}~\bibnamefont {Zoller}},\ }\href {\doibase 10.1103/PhysRevLett.87.037901}
  {\bibfield  {journal} {\bibinfo  {journal} {Phys. Rev. Lett.}\ }\textbf
  {\bibinfo {volume} {87}},\ \bibinfo {pages} {037901} (\bibinfo {year}
  {2001})}\BibitemShut {NoStop}%
\bibitem [{\citenamefont {Comparat}\ and\ \citenamefont
  {Pillet}(2010)}]{Comparat2010}%
  \BibitemOpen
  \bibfield  {author} {\bibinfo {author} {\bibfnamefont {D.}~\bibnamefont
  {Comparat}}\ and\ \bibinfo {author} {\bibfnamefont {P.}~\bibnamefont
  {Pillet}},\ }\href {\doibase 10.1364/JOSAB.27.00A208} {\bibfield  {journal}
  {\bibinfo  {journal} {J. Opt. Soc. Am. B}\ }\textbf {\bibinfo {volume}
  {27}},\ \bibinfo {pages} {A208} (\bibinfo {year} {2010})}\BibitemShut
  {NoStop}%
\bibitem [{\citenamefont {Stanojevic}\ and\ \citenamefont
  {C\^ot\'e}(2009)}]{Stanojevic2009}%
  \BibitemOpen
  \bibfield  {author} {\bibinfo {author} {\bibfnamefont {J.}~\bibnamefont
  {Stanojevic}}\ and\ \bibinfo {author} {\bibfnamefont {R.}~\bibnamefont
  {C\^ot\'e}},\ }\href {\doibase 10.1103/PhysRevA.80.033418} {\bibfield
  {journal} {\bibinfo  {journal} {Phys. Rev. A}\ }\textbf {\bibinfo {volume}
  {80}},\ \bibinfo {pages} {033418} (\bibinfo {year} {2009})}\BibitemShut
  {NoStop}%
\bibitem [{\citenamefont {G\"arttner}\ \emph {et~al.}(2012)\citenamefont
  {G\"arttner}, \citenamefont {Heeg}, \citenamefont {Gasenzer},\ and\
  \citenamefont {Evers}}]{Garttner2012}%
  \BibitemOpen
  \bibfield  {author} {\bibinfo {author} {\bibfnamefont {M.}~\bibnamefont
  {G\"arttner}}, \bibinfo {author} {\bibfnamefont {K.~P.}\ \bibnamefont
  {Heeg}}, \bibinfo {author} {\bibfnamefont {T.}~\bibnamefont {Gasenzer}}, \
  and\ \bibinfo {author} {\bibfnamefont {J.}~\bibnamefont {Evers}},\ }\href
  {\doibase 10.1103/PhysRevA.86.033422} {\bibfield  {journal} {\bibinfo
  {journal} {Phys. Rev. A}\ }\textbf {\bibinfo {volume} {86}},\ \bibinfo
  {pages} {033422} (\bibinfo {year} {2012})}\BibitemShut {NoStop}%
\bibitem [{\citenamefont {Dudin}\ and\ \citenamefont
  {Kuzmich}(2012)}]{Dudin2012}%
  \BibitemOpen
  \bibfield  {author} {\bibinfo {author} {\bibfnamefont {Y.~O.}\ \bibnamefont
  {Dudin}}\ and\ \bibinfo {author} {\bibfnamefont {A.}~\bibnamefont
  {Kuzmich}},\ }\href {\doibase DOI: 10.1038/NPHYS2413} {\bibfield  {journal}
  {\bibinfo  {journal} {Science}\ }\textbf {\bibinfo {volume} {336}},\ \bibinfo
  {pages} {887} (\bibinfo {year} {2012})}\BibitemShut {NoStop}%
\bibitem [{\citenamefont {Guerlin}\ \emph {et~al.}(2010)\citenamefont
  {Guerlin}, \citenamefont {Brion}, \citenamefont {Esslinger},\ and\
  \citenamefont {M\o{}lmer}}]{Guerlin2010}%
  \BibitemOpen
  \bibfield  {author} {\bibinfo {author} {\bibfnamefont {C.}~\bibnamefont
  {Guerlin}}, \bibinfo {author} {\bibfnamefont {E.}~\bibnamefont {Brion}},
  \bibinfo {author} {\bibfnamefont {T.}~\bibnamefont {Esslinger}}, \ and\
  \bibinfo {author} {\bibfnamefont {K.}~\bibnamefont {M\o{}lmer}},\ }\href
  {\doibase 10.1103/PhysRevA.82.053832} {\bibfield  {journal} {\bibinfo
  {journal} {Phys. Rev. A}\ }\textbf {\bibinfo {volume} {82}},\ \bibinfo
  {pages} {053832} (\bibinfo {year} {2010})}\BibitemShut {NoStop}%
\bibitem [{\citenamefont {Firstenberg}\ \emph
  {et~al.}(2013{\natexlab{a}})\citenamefont {Firstenberg}, \citenamefont
  {Peyronel}, \citenamefont {Liang}, \citenamefont {Gorshkov}, \citenamefont
  {Lukin},\ and\ \citenamefont {Vuletic}}]{Firstenberg2013}%
  \BibitemOpen
  \bibfield  {author} {\bibinfo {author} {\bibfnamefont {O.}~\bibnamefont
  {Firstenberg}}, \bibinfo {author} {\bibfnamefont {T.}~\bibnamefont
  {Peyronel}}, \bibinfo {author} {\bibfnamefont {Q.-Y.}\ \bibnamefont {Liang}},
  \bibinfo {author} {\bibfnamefont {A.~V.}\ \bibnamefont {Gorshkov}}, \bibinfo
  {author} {\bibfnamefont {M.~D.}\ \bibnamefont {Lukin}}, \ and\ \bibinfo
  {author} {\bibfnamefont {V.}~\bibnamefont {Vuletic}},\ }\href {\doibase
  doi:10.1038/nature12512} {\bibfield  {journal} {\bibinfo  {journal} {Nature}\
  }\textbf {\bibinfo {volume} {502}},\ \bibinfo {pages} {71} (\bibinfo {year}
  {2013}{\natexlab{a}})}\BibitemShut {NoStop}%
\bibitem [{\citenamefont {Firstenberg}\ \emph
  {et~al.}(2013{\natexlab{b}})\citenamefont {Firstenberg}, \citenamefont
  {Lukin}, \citenamefont {Peyronel}, \citenamefont {Liang}, \citenamefont
  {Vuletic}, \citenamefont {Gorshkov}, \citenamefont {Hofferberth},\ and\
  \citenamefont {Pohl}}]{Firstenberg2013a}%
  \BibitemOpen
  \bibfield  {author} {\bibinfo {author} {\bibfnamefont {O.}~\bibnamefont
  {Firstenberg}}, \bibinfo {author} {\bibfnamefont {M.}~\bibnamefont {Lukin}},
  \bibinfo {author} {\bibfnamefont {T.}~\bibnamefont {Peyronel}}, \bibinfo
  {author} {\bibfnamefont {Q.-Y.}\ \bibnamefont {Liang}}, \bibinfo {author}
  {\bibfnamefont {V.}~\bibnamefont {Vuletic}}, \bibinfo {author} {\bibfnamefont
  {A.}~\bibnamefont {Gorshkov}}, \bibinfo {author} {\bibfnamefont
  {S.}~\bibnamefont {Hofferberth}}, \ and\ \bibinfo {author} {\bibfnamefont
  {T.}~\bibnamefont {Pohl}},\ }\href {\doibase 10.1364/OPN.24.12.000048}
  {\bibfield  {journal} {\bibinfo  {journal} {Optics and Photonics News}\
  }\textbf {\bibinfo {volume} {24}},\ \bibinfo {pages} {48} (\bibinfo {year}
  {2013}{\natexlab{b}})}\BibitemShut {NoStop}%
\bibitem [{\citenamefont {Grimm}\ \emph {et~al.}(2000)\citenamefont {Grimm},
  \citenamefont {Weidem\"uller},\ and\ \citenamefont {Ovchnnikov}}]{Grimm2000}%
  \BibitemOpen
  \bibfield  {author} {\bibinfo {author} {\bibfnamefont {R.}~\bibnamefont
  {Grimm}}, \bibinfo {author} {\bibfnamefont {M.}~\bibnamefont
  {Weidem\"uller}}, \ and\ \bibinfo {author} {\bibfnamefont {Y.~B.}\
  \bibnamefont {Ovchnnikov}},\ }\href {\doibase 10.1016/S1049-250X(08)60186-X}
  {\bibfield  {journal} {\bibinfo  {journal} {Adv. At., Mol., Opt. Phys.}\
  }\textbf {\bibinfo {volume} {42}},\ \bibinfo {pages} {95} (\bibinfo {year}
  {2000})}\BibitemShut {NoStop}%
\bibitem [{\citenamefont {Saffman}\ \emph {et~al.}(2010)\citenamefont
  {Saffman}, \citenamefont {Walker},\ and\ \citenamefont
  {M\o{}lmer}}]{Saffman2010}%
  \BibitemOpen
  \bibfield  {author} {\bibinfo {author} {\bibfnamefont {M.}~\bibnamefont
  {Saffman}}, \bibinfo {author} {\bibfnamefont {T.~G.}\ \bibnamefont {Walker}},
  \ and\ \bibinfo {author} {\bibfnamefont {K.}~\bibnamefont {M\o{}lmer}},\
  }\href {\doibase 10.1103/RevModPhys.82.2313} {\bibfield  {journal} {\bibinfo
  {journal} {Rev. Mod. Phys.}\ }\textbf {\bibinfo {volume} {82}},\ \bibinfo
  {pages} {2313} (\bibinfo {year} {2010})}\BibitemShut {NoStop}%
\bibitem [{\citenamefont {Beterov}\ \emph {et~al.}(2011)\citenamefont
  {Beterov}, \citenamefont {Tretyakov}, \citenamefont {Entin}, \citenamefont
  {Yakshina}, \citenamefont {Ryabtsev}, \citenamefont {MacCormick},\ and\
  \citenamefont {Bergamini}}]{Beterov2011}%
  \BibitemOpen
  \bibfield  {author} {\bibinfo {author} {\bibfnamefont {I.~I.}\ \bibnamefont
  {Beterov}}, \bibinfo {author} {\bibfnamefont {D.~B.}\ \bibnamefont
  {Tretyakov}}, \bibinfo {author} {\bibfnamefont {V.~M.}\ \bibnamefont
  {Entin}}, \bibinfo {author} {\bibfnamefont {E.~A.}\ \bibnamefont {Yakshina}},
  \bibinfo {author} {\bibfnamefont {I.~I.}\ \bibnamefont {Ryabtsev}}, \bibinfo
  {author} {\bibfnamefont {C.}~\bibnamefont {MacCormick}}, \ and\ \bibinfo
  {author} {\bibfnamefont {S.}~\bibnamefont {Bergamini}},\ }\href {\doibase
  10.1103/PhysRevA.84.023413} {\bibfield  {journal} {\bibinfo  {journal} {Phys.
  Rev. A}\ }\textbf {\bibinfo {volume} {84}},\ \bibinfo {pages} {023413}
  (\bibinfo {year} {2011})}\BibitemShut {NoStop}%
\bibitem [{\citenamefont {Beterov}\ \emph {et~al.}(2013)\citenamefont
  {Beterov}, \citenamefont {Saffman}, \citenamefont {Yakshina}, \citenamefont
  {Zhukov}, \citenamefont {Tretyakov}, \citenamefont {Entin}, \citenamefont
  {Ryabtsev}, \citenamefont {Mansell}, \citenamefont {MacCormick},
  \citenamefont {Bergamini},\ and\ \citenamefont {Fedoruk}}]{Beterov2013}%
  \BibitemOpen
  \bibfield  {author} {\bibinfo {author} {\bibfnamefont {I.~I.}\ \bibnamefont
  {Beterov}}, \bibinfo {author} {\bibfnamefont {M.}~\bibnamefont {Saffman}},
  \bibinfo {author} {\bibfnamefont {E.~A.}\ \bibnamefont {Yakshina}}, \bibinfo
  {author} {\bibfnamefont {V.~P.}\ \bibnamefont {Zhukov}}, \bibinfo {author}
  {\bibfnamefont {D.~B.}\ \bibnamefont {Tretyakov}}, \bibinfo {author}
  {\bibfnamefont {V.~M.}\ \bibnamefont {Entin}}, \bibinfo {author}
  {\bibfnamefont {I.~I.}\ \bibnamefont {Ryabtsev}}, \bibinfo {author}
  {\bibfnamefont {C.~W.}\ \bibnamefont {Mansell}}, \bibinfo {author}
  {\bibfnamefont {C.}~\bibnamefont {MacCormick}}, \bibinfo {author}
  {\bibfnamefont {S.}~\bibnamefont {Bergamini}}, \ and\ \bibinfo {author}
  {\bibfnamefont {M.~P.}\ \bibnamefont {Fedoruk}},\ }\href {\doibase
  10.1103/PhysRevA.88.010303} {\bibfield  {journal} {\bibinfo  {journal} {Phys.
  Rev. A}\ }\textbf {\bibinfo {volume} {88}},\ \bibinfo {pages} {010303(R)}
  (\bibinfo {year} {2013})}\BibitemShut {NoStop}%
\bibitem [{\citenamefont {Beterov}\ \emph {et~al.}(2014)\citenamefont
  {Beterov}, \citenamefont {Saffman}, \citenamefont {Yakshina}, \citenamefont
  {Zhukov}, \citenamefont {Tretyakov}, \citenamefont {Entin}, \citenamefont
  {Ryabtsev}, \citenamefont {Mansell}, \citenamefont {MacCormick},
  \citenamefont {Bergamini},\ and\ \citenamefont {Fedoruk}}]{Beterov2014}%
  \BibitemOpen
  \bibfield  {author} {\bibinfo {author} {\bibfnamefont {I.~I.}\ \bibnamefont
  {Beterov}}, \bibinfo {author} {\bibfnamefont {M.}~\bibnamefont {Saffman}},
  \bibinfo {author} {\bibfnamefont {E.~A.}\ \bibnamefont {Yakshina}}, \bibinfo
  {author} {\bibfnamefont {V.~P.}\ \bibnamefont {Zhukov}}, \bibinfo {author}
  {\bibfnamefont {D.~B.}\ \bibnamefont {Tretyakov}}, \bibinfo {author}
  {\bibfnamefont {V.~M.}\ \bibnamefont {Entin}}, \bibinfo {author}
  {\bibfnamefont {I.~I.}\ \bibnamefont {Ryabtsev}}, \bibinfo {author}
  {\bibfnamefont {C.~W.}\ \bibnamefont {Mansell}}, \bibinfo {author}
  {\bibfnamefont {C.}~\bibnamefont {MacCormick}}, \bibinfo {author}
  {\bibfnamefont {S.}~\bibnamefont {Bergamini}}, \ and\ \bibinfo {author}
  {\bibfnamefont {M.~P.}\ \bibnamefont {Fedoruk}},\ }\href {\doibase
  doi:10.1088/1054-660X/24/7/074013} {\bibfield  {journal} {\bibinfo  {journal}
  {Laser Physics}\ }\textbf {\bibinfo {volume} {24}},\ \bibinfo {pages}
  {074013} (\bibinfo {year} {2014})}\BibitemShut {NoStop}%
\bibitem [{\citenamefont {Isenhower}\ \emph {et~al.}(2010)\citenamefont
  {Isenhower}, \citenamefont {Urban}, \citenamefont {Zhang}, \citenamefont
  {Gill}, \citenamefont {Henage}, \citenamefont {Johnson}, \citenamefont
  {Walker},\ and\ \citenamefont {Saffman}}]{Isenhower2010}%
  \BibitemOpen
  \bibfield  {author} {\bibinfo {author} {\bibfnamefont {L.}~\bibnamefont
  {Isenhower}}, \bibinfo {author} {\bibfnamefont {E.}~\bibnamefont {Urban}},
  \bibinfo {author} {\bibfnamefont {X.~L.}\ \bibnamefont {Zhang}}, \bibinfo
  {author} {\bibfnamefont {A.~T.}\ \bibnamefont {Gill}}, \bibinfo {author}
  {\bibfnamefont {T.}~\bibnamefont {Henage}}, \bibinfo {author} {\bibfnamefont
  {T.~A.}\ \bibnamefont {Johnson}}, \bibinfo {author} {\bibfnamefont {T.~G.}\
  \bibnamefont {Walker}}, \ and\ \bibinfo {author} {\bibfnamefont
  {M.}~\bibnamefont {Saffman}},\ }\href {\doibase
  10.1103/PhysRevLett.104.010503} {\bibfield  {journal} {\bibinfo  {journal}
  {Phys. Rev. Lett.}\ }\textbf {\bibinfo {volume} {104}},\ \bibinfo {pages}
  {010503} (\bibinfo {year} {2010})}\BibitemShut {NoStop}%
\bibitem [{\citenamefont {Wilk}\ \emph {et~al.}(2010)\citenamefont {Wilk},
  \citenamefont {Ga\"etan}, \citenamefont {Evellin}, \citenamefont {Wolters},
  \citenamefont {Miroshnychenko}, \citenamefont {Grangier},\ and\ \citenamefont
  {Browaeys}}]{Wilk2010}%
  \BibitemOpen
  \bibfield  {author} {\bibinfo {author} {\bibfnamefont {T.}~\bibnamefont
  {Wilk}}, \bibinfo {author} {\bibfnamefont {A.}~\bibnamefont {Ga\"etan}},
  \bibinfo {author} {\bibfnamefont {C.}~\bibnamefont {Evellin}}, \bibinfo
  {author} {\bibfnamefont {J.}~\bibnamefont {Wolters}}, \bibinfo {author}
  {\bibfnamefont {Y.}~\bibnamefont {Miroshnychenko}}, \bibinfo {author}
  {\bibfnamefont {P.}~\bibnamefont {Grangier}}, \ and\ \bibinfo {author}
  {\bibfnamefont {A.}~\bibnamefont {Browaeys}},\ }\href {\doibase
  10.1103/PhysRevLett.104.010502} {\bibfield  {journal} {\bibinfo  {journal}
  {Phys. Rev. Lett.}\ }\textbf {\bibinfo {volume} {104}},\ \bibinfo {pages}
  {010502} (\bibinfo {year} {2010})}\BibitemShut {NoStop}%
\bibitem [{\citenamefont {Urban}\ \emph {et~al.}(2009)\citenamefont {Urban},
  \citenamefont {Johnson}, \citenamefont {Henage}, \citenamefont {Isenhower},
  \citenamefont {Yavuz}, \citenamefont {Walker},\ and\ \citenamefont
  {Saffman}}]{Urban2009}%
  \BibitemOpen
  \bibfield  {author} {\bibinfo {author} {\bibfnamefont {E.}~\bibnamefont
  {Urban}}, \bibinfo {author} {\bibfnamefont {T.~A.}\ \bibnamefont {Johnson}},
  \bibinfo {author} {\bibfnamefont {T.}~\bibnamefont {Henage}}, \bibinfo
  {author} {\bibfnamefont {L.}~\bibnamefont {Isenhower}}, \bibinfo {author}
  {\bibfnamefont {D.~D.}\ \bibnamefont {Yavuz}}, \bibinfo {author}
  {\bibfnamefont {T.~G.}\ \bibnamefont {Walker}}, \ and\ \bibinfo {author}
  {\bibfnamefont {M.}~\bibnamefont {Saffman}},\ }\href {\doibase DOI:
  10.1038/NPHYS1178} {\bibfield  {journal} {\bibinfo  {journal} {Nature
  Physics}\ }\textbf {\bibinfo {volume} {5}},\ \bibinfo {pages} {110} (\bibinfo
  {year} {2009})}\BibitemShut {NoStop}%
\bibitem [{\citenamefont {Nogrette}\ \emph {et~al.}(2014)\citenamefont
  {Nogrette}, \citenamefont {Labuhn}, \citenamefont {Ravets}, \citenamefont
  {Barredo}, \citenamefont {Beguin}, \citenamefont {Vernier}, \citenamefont
  {Lahaye},\ and\ \citenamefont {Browaeys}}]{Nogrette2014}%
  \BibitemOpen
  \bibfield  {author} {\bibinfo {author} {\bibfnamefont {F.}~\bibnamefont
  {Nogrette}}, \bibinfo {author} {\bibfnamefont {H.}~\bibnamefont {Labuhn}},
  \bibinfo {author} {\bibfnamefont {S.}~\bibnamefont {Ravets}}, \bibinfo
  {author} {\bibfnamefont {D.}~\bibnamefont {Barredo}}, \bibinfo {author}
  {\bibfnamefont {L.}~\bibnamefont {Beguin}}, \bibinfo {author} {\bibfnamefont
  {A.}~\bibnamefont {Vernier}}, \bibinfo {author} {\bibfnamefont
  {T.}~\bibnamefont {Lahaye}}, \ and\ \bibinfo {author} {\bibfnamefont
  {A.}~\bibnamefont {Browaeys}},\ }\href {\doibase 10.1103/PhysRevX.4.021034}
  {\bibfield  {journal} {\bibinfo  {journal} {Phys. Rev. X}\ }\textbf {\bibinfo
  {volume} {4}},\ \bibinfo {pages} {021034} (\bibinfo {year}
  {2014})}\BibitemShut {NoStop}%
\bibitem [{\citenamefont {Hume}\ \emph {et~al.}(2013)\citenamefont {Hume},
  \citenamefont {Stroescu}, \citenamefont {Joos}, \citenamefont {Muessel},
  \citenamefont {Strobel},\ and\ \citenamefont {Oberthaler}}]{Hume2013}%
  \BibitemOpen
  \bibfield  {author} {\bibinfo {author} {\bibfnamefont {D.~B.}\ \bibnamefont
  {Hume}}, \bibinfo {author} {\bibfnamefont {I.}~\bibnamefont {Stroescu}},
  \bibinfo {author} {\bibfnamefont {M.}~\bibnamefont {Joos}}, \bibinfo {author}
  {\bibfnamefont {W.}~\bibnamefont {Muessel}}, \bibinfo {author} {\bibfnamefont
  {H.}~\bibnamefont {Strobel}}, \ and\ \bibinfo {author} {\bibfnamefont
  {M.~K.}\ \bibnamefont {Oberthaler}},\ }\href {\doibase
  10.1103/PhysRevLett.111.253001} {\bibfield  {journal} {\bibinfo  {journal}
  {Phys. Rev. Lett.}\ }\textbf {\bibinfo {volume} {111}},\ \bibinfo {pages}
  {253001} (\bibinfo {year} {2013})}\BibitemShut {NoStop}%
\bibitem [{\citenamefont {Stebbings}\ \emph {et~al.}(1975)\citenamefont
  {Stebbings}, \citenamefont {Latimer}, \citenamefont {West}, \citenamefont
  {Dunning},\ and\ \citenamefont {Cook}}]{Stebbings1975}%
  \BibitemOpen
  \bibfield  {author} {\bibinfo {author} {\bibfnamefont {R.~F.}\ \bibnamefont
  {Stebbings}}, \bibinfo {author} {\bibfnamefont {C.~J.}\ \bibnamefont
  {Latimer}}, \bibinfo {author} {\bibfnamefont {W.~P.}\ \bibnamefont {West}},
  \bibinfo {author} {\bibfnamefont {F.~B.}\ \bibnamefont {Dunning}}, \ and\
  \bibinfo {author} {\bibfnamefont {T.~B.}\ \bibnamefont {Cook}},\ }\href
  {\doibase 10.1103/PhysRevA.12.1453} {\bibfield  {journal} {\bibinfo
  {journal} {Physical Review}\ }\textbf {\bibinfo {volume} {12}},\ \bibinfo
  {pages} {1453} (\bibinfo {year} {1975})}\BibitemShut {NoStop}%
\bibitem [{\citenamefont {Ryabtsev}\ \emph {et~al.}(2007)\citenamefont
  {Ryabtsev}, \citenamefont {Tretyakov}, \citenamefont {Beterov},\ and\
  \citenamefont {Entin}}]{Ryabtsev2007}%
  \BibitemOpen
  \bibfield  {author} {\bibinfo {author} {\bibfnamefont {I.~I.}\ \bibnamefont
  {Ryabtsev}}, \bibinfo {author} {\bibfnamefont {D.~B.}\ \bibnamefont
  {Tretyakov}}, \bibinfo {author} {\bibfnamefont {I.~I.}\ \bibnamefont
  {Beterov}}, \ and\ \bibinfo {author} {\bibfnamefont {V.~M.}\ \bibnamefont
  {Entin}},\ }\href {\doibase 10.1103/PhysRevA.76.012722} {\bibfield  {journal}
  {\bibinfo  {journal} {Phys. Rev. A}\ }\textbf {\bibinfo {volume} {76}},\
  \bibinfo {pages} {012722} (\bibinfo {year} {2007})}\BibitemShut {NoStop}%
\bibitem [{\citenamefont {Ryabtsev}\ \emph
  {et~al.}(2010{\natexlab{a}})\citenamefont {Ryabtsev}, \citenamefont
  {Tretyakov}, \citenamefont {Beterov},\ and\ \citenamefont
  {Entin}}]{Ryabtsev2010}%
  \BibitemOpen
  \bibfield  {author} {\bibinfo {author} {\bibfnamefont {I.~I.}\ \bibnamefont
  {Ryabtsev}}, \bibinfo {author} {\bibfnamefont {D.~B.}\ \bibnamefont
  {Tretyakov}}, \bibinfo {author} {\bibfnamefont {I.~I.}\ \bibnamefont
  {Beterov}}, \ and\ \bibinfo {author} {\bibfnamefont {V.~M.}\ \bibnamefont
  {Entin}},\ }\href {\doibase 10.1103/PhysRevLett.104.073003} {\bibfield
  {journal} {\bibinfo  {journal} {Phys. Rev. Lett.}\ }\textbf {\bibinfo
  {volume} {104}},\ \bibinfo {pages} {073003} (\bibinfo {year}
  {2010}{\natexlab{a}})}\BibitemShut {NoStop}%
\bibitem [{\citenamefont {Loudon}(1983)}]{Loudon1983}%
  \BibitemOpen
  \bibfield  {author} {\bibinfo {author} {\bibfnamefont {R.}~\bibnamefont
  {Loudon}},\ }\href@noop {} {\emph {\bibinfo {title} {The quantum theory of
  light}}}\ (\bibinfo  {publisher} {Oxford University Press, Oxford},\ \bibinfo
  {year} {1983})\BibitemShut {NoStop}%
\bibitem [{\citenamefont {Brune}\ \emph {et~al.}(1996)\citenamefont {Brune},
  \citenamefont {Schmidt-Kaler}, \citenamefont {Maali}, \citenamefont {Dreyer},
  \citenamefont {Hagley}, \citenamefont {Raimond},\ and\ \citenamefont
  {Haroche}}]{Brune1996}%
  \BibitemOpen
  \bibfield  {author} {\bibinfo {author} {\bibfnamefont {M.}~\bibnamefont
  {Brune}}, \bibinfo {author} {\bibfnamefont {F.}~\bibnamefont
  {Schmidt-Kaler}}, \bibinfo {author} {\bibfnamefont {A.}~\bibnamefont
  {Maali}}, \bibinfo {author} {\bibfnamefont {J.}~\bibnamefont {Dreyer}},
  \bibinfo {author} {\bibfnamefont {E.}~\bibnamefont {Hagley}}, \bibinfo
  {author} {\bibfnamefont {J.~M.}\ \bibnamefont {Raimond}}, \ and\ \bibinfo
  {author} {\bibfnamefont {S.}~\bibnamefont {Haroche}},\ }\href {\doibase
  10.1103/PhysRevLett.76.1800} {\bibfield  {journal} {\bibinfo  {journal}
  {Phys. Rev. Lett.}\ }\textbf {\bibinfo {volume} {76}},\ \bibinfo {pages}
  {1800} (\bibinfo {year} {1996})}\BibitemShut {NoStop}%
\bibitem [{\citenamefont {Gea-Banacloche}(1990)}]{Gea-Banacloche1990}%
  \BibitemOpen
  \bibfield  {author} {\bibinfo {author} {\bibfnamefont {J.}~\bibnamefont
  {Gea-Banacloche}},\ }\href {\doibase 10.1103/PhysRevLett.65.3385} {\bibfield
  {journal} {\bibinfo  {journal} {Phys. Rev. Lett.}\ }\textbf {\bibinfo
  {volume} {65}},\ \bibinfo {pages} {3385} (\bibinfo {year}
  {1990})}\BibitemShut {NoStop}%
\bibitem [{\citenamefont {Louisell}(1973)}]{Louisell1973}%
  \BibitemOpen
  \bibfield  {author} {\bibinfo {author} {\bibfnamefont {W.}~\bibnamefont
  {Louisell}},\ }\href@noop {} {\emph {\bibinfo {title} {Quantum Statistical
  Properties of Radiation}}}\ (\bibinfo  {publisher} {Wiley, New York},\
  \bibinfo {year} {1973})\BibitemShut {NoStop}%
\bibitem [{\citenamefont {Cohen-Tannoudji}(2004)}]{Cohen-Tannoudji2004}%
  \BibitemOpen
  \bibfield  {author} {\bibinfo {author} {\bibfnamefont {C.}~\bibnamefont
  {Cohen-Tannoudji}},\ }\href@noop {} {\emph {\bibinfo {title} {Atoms in
  Electromagnetic Fields. Second Edition}}}\ (\bibinfo  {publisher} {World
  Scientific, Singapore},\ \bibinfo {year} {2004})\BibitemShut {NoStop}%
\bibitem [{\citenamefont {M\o{}lmer}(1997)}]{Molmer1997}%
  \BibitemOpen
  \bibfield  {author} {\bibinfo {author} {\bibfnamefont {K.}~\bibnamefont
  {M\o{}lmer}},\ }\href {\doibase 10.1103/PhysRevA.55.3195} {\bibfield
  {journal} {\bibinfo  {journal} {Phys. Rev. A}\ }\textbf {\bibinfo {volume}
  {55}},\ \bibinfo {pages} {3195} (\bibinfo {year} {1997})}\BibitemShut
  {NoStop}%
\bibitem [{\citenamefont {Honer}\ \emph {et~al.}(2011)\citenamefont {Honer},
  \citenamefont {L\"ow}, \citenamefont {Weimer}, \citenamefont {Pfau},\ and\
  \citenamefont {B\"uchler}}]{Honer2011}%
  \BibitemOpen
  \bibfield  {author} {\bibinfo {author} {\bibfnamefont {J.}~\bibnamefont
  {Honer}}, \bibinfo {author} {\bibfnamefont {R.}~\bibnamefont {L\"ow}},
  \bibinfo {author} {\bibfnamefont {H.}~\bibnamefont {Weimer}}, \bibinfo
  {author} {\bibfnamefont {T.}~\bibnamefont {Pfau}}, \ and\ \bibinfo {author}
  {\bibfnamefont {H.~P.}\ \bibnamefont {B\"uchler}},\ }\href {\doibase
  10.1103/PhysRevLett.107.093601} {\bibfield  {journal} {\bibinfo  {journal}
  {Phys. Rev. Lett.}\ }\textbf {\bibinfo {volume} {107}},\ \bibinfo {pages}
  {093601} (\bibinfo {year} {2011})}\BibitemShut {NoStop}%
\bibitem [{\citenamefont {Ryabtsev}\ \emph
  {et~al.}(2010{\natexlab{b}})\citenamefont {Ryabtsev}, \citenamefont
  {Tretyakov}, \citenamefont {Beterov}, \citenamefont {Entin},\ and\
  \citenamefont {Yakshina}}]{Ryabtsev2010a}%
  \BibitemOpen
  \bibfield  {author} {\bibinfo {author} {\bibfnamefont {I.~I.}\ \bibnamefont
  {Ryabtsev}}, \bibinfo {author} {\bibfnamefont {D.~B.}\ \bibnamefont
  {Tretyakov}}, \bibinfo {author} {\bibfnamefont {I.~I.}\ \bibnamefont
  {Beterov}}, \bibinfo {author} {\bibfnamefont {V.~M.}\ \bibnamefont {Entin}},
  \ and\ \bibinfo {author} {\bibfnamefont {E.~A.}\ \bibnamefont {Yakshina}},\
  }\href {\doibase 10.1103/PhysRevA.82.053409} {\bibfield  {journal} {\bibinfo
  {journal} {Phys. Rev. A}\ }\textbf {\bibinfo {volume} {82}},\ \bibinfo
  {pages} {053409} (\bibinfo {year} {2010}{\natexlab{b}})}\BibitemShut
  {NoStop}%
\bibitem [{\citenamefont {Beterov}\ \emph {et~al.}(2009)\citenamefont
  {Beterov}, \citenamefont {Ryabtsev}, \citenamefont {Tretyakov},\ and\
  \citenamefont {Entin}}]{Beterov2009}%
  \BibitemOpen
  \bibfield  {author} {\bibinfo {author} {\bibfnamefont {I.~I.}\ \bibnamefont
  {Beterov}}, \bibinfo {author} {\bibfnamefont {I.~I.}\ \bibnamefont
  {Ryabtsev}}, \bibinfo {author} {\bibfnamefont {D.~B.}\ \bibnamefont
  {Tretyakov}}, \ and\ \bibinfo {author} {\bibfnamefont {V.~M.}\ \bibnamefont
  {Entin}},\ }\href {\doibase 10.1103/PhysRevA.79.052504} {\bibfield  {journal}
  {\bibinfo  {journal} {Phys. Rev. A}\ }\textbf {\bibinfo {volume} {79}},\
  \bibinfo {pages} {052504} (\bibinfo {year} {2009})}\BibitemShut {NoStop}%
\bibitem [{\citenamefont {Petrosyan}\ and\ \citenamefont
  {M\o{}lmer}(2013)}]{Petrosyan2013}%
  \BibitemOpen
  \bibfield  {author} {\bibinfo {author} {\bibfnamefont {D.}~\bibnamefont
  {Petrosyan}}\ and\ \bibinfo {author} {\bibfnamefont {K.}~\bibnamefont
  {M\o{}lmer}},\ }\href {\doibase 10.1103/PhysRevA.87.033416} {\bibfield
  {journal} {\bibinfo  {journal} {Phys. Rev. A}\ }\textbf {\bibinfo {volume}
  {87}},\ \bibinfo {pages} {033416} (\bibinfo {year} {2013})}\BibitemShut
  {NoStop}%
\bibitem [{\citenamefont {Schlosser}\ \emph {et~al.}(2002)\citenamefont
  {Schlosser}, \citenamefont {Reymond},\ and\ \citenamefont
  {Grangier}}]{Schlosser2002}%
  \BibitemOpen
  \bibfield  {author} {\bibinfo {author} {\bibfnamefont {N.}~\bibnamefont
  {Schlosser}}, \bibinfo {author} {\bibfnamefont {G.}~\bibnamefont {Reymond}},
  \ and\ \bibinfo {author} {\bibfnamefont {P.}~\bibnamefont {Grangier}},\
  }\href {\doibase 10.1103/PhysRevLett.89.023005} {\bibfield  {journal}
  {\bibinfo  {journal} {Phys. Rev. Lett.}\ }\textbf {\bibinfo {volume} {89}},\
  \bibinfo {pages} {023005} (\bibinfo {year} {2002})}\BibitemShut {NoStop}%
\bibitem [{\citenamefont {Reymond}\ \emph {et~al.}(2003)\citenamefont
  {Reymond}, \citenamefont {Schlosser}, \citenamefont {Protsenko},\ and\
  \citenamefont {Grangier}}]{Reymond2003}%
  \BibitemOpen
  \bibfield  {author} {\bibinfo {author} {\bibfnamefont {G.}~\bibnamefont
  {Reymond}}, \bibinfo {author} {\bibfnamefont {N.}~\bibnamefont {Schlosser}},
  \bibinfo {author} {\bibfnamefont {I.}~\bibnamefont {Protsenko}}, \ and\
  \bibinfo {author} {\bibfnamefont {P.}~\bibnamefont {Grangier}},\ }\href
  {\doibase 10.1098/rsta.2003.1219} {\bibfield  {journal} {\bibinfo  {journal}
  {Philos. Trans. Roy. Soc. London A}\ }\textbf {\bibinfo {volume} {361}},\
  \bibinfo {pages} {1527} (\bibinfo {year} {2003})}\BibitemShut {NoStop}%
\bibitem [{\citenamefont {Mansell}\ and\ \citenamefont
  {Bergamini}(2014)}]{Mansell2014}%
  \BibitemOpen
  \bibfield  {author} {\bibinfo {author} {\bibfnamefont {C.~W.}\ \bibnamefont
  {Mansell}}\ and\ \bibinfo {author} {\bibfnamefont {S.}~\bibnamefont
  {Bergamini}},\ }\href {\doibase 10.1088/1367-2630/16/5/053045} {\bibfield
  {journal} {\bibinfo  {journal} {New J. Phys.}\ }\textbf {\bibinfo {volume}
  {16}},\ \bibinfo {pages} {053045} (\bibinfo {year} {2014})}\BibitemShut
  {NoStop}%
\bibitem [{\citenamefont {Akkermans}\ \emph {et~al.}(2008)\citenamefont
  {Akkermans}, \citenamefont {Gero},\ and\ \citenamefont
  {Kaiser}}]{Akkermans2008}%
  \BibitemOpen
  \bibfield  {author} {\bibinfo {author} {\bibfnamefont {E.}~\bibnamefont
  {Akkermans}}, \bibinfo {author} {\bibfnamefont {A.}~\bibnamefont {Gero}}, \
  and\ \bibinfo {author} {\bibfnamefont {R.}~\bibnamefont {Kaiser}},\ }\href
  {\doibase 10.1103/PhysRevLett.101.103602} {\bibfield  {journal} {\bibinfo
  {journal} {Phys. Rev. Lett.}\ }\textbf {\bibinfo {volume} {101}},\ \bibinfo
  {pages} {103602} (\bibinfo {year} {2008})}\BibitemShut {NoStop}%
\bibitem [{\citenamefont {Garraway}(2011)}]{Garraway2011}%
  \BibitemOpen
  \bibfield  {author} {\bibinfo {author} {\bibfnamefont {B.~M.}\ \bibnamefont
  {Garraway}},\ }\href {\doibase 10.1098/rsta.2010.0333} {\bibfield  {journal}
  {\bibinfo  {journal} {Philos. Trans. Roy. Soc. London A}\ }\textbf {\bibinfo
  {volume} {369}},\ \bibinfo {pages} {1137} (\bibinfo {year}
  {2011})}\BibitemShut {NoStop}%
\bibitem [{\citenamefont {Bienaime}\ \emph {et~al.}(2012)\citenamefont
  {Bienaime}, \citenamefont {Piovella},\ and\ \citenamefont
  {Kaiser}}]{Bienaime2012}%
  \BibitemOpen
  \bibfield  {author} {\bibinfo {author} {\bibfnamefont {T.}~\bibnamefont
  {Bienaime}}, \bibinfo {author} {\bibfnamefont {N.}~\bibnamefont {Piovella}},
  \ and\ \bibinfo {author} {\bibfnamefont {R.}~\bibnamefont {Kaiser}},\ }\href
  {\doibase 10.1103/PhysRevLett.108.123602} {\bibfield  {journal} {\bibinfo
  {journal} {Phys. Rev. Lett.}\ }\textbf {\bibinfo {volume} {108}},\ \bibinfo
  {pages} {123602} (\bibinfo {year} {2012})}\BibitemShut {NoStop}%
\bibitem [{\citenamefont {Maxwell}\ \emph {et~al.}(2013)\citenamefont
  {Maxwell}, \citenamefont {Szwer}, \citenamefont {Paredes-Barato},
  \citenamefont {Busche}, \citenamefont {Pritchard}, \citenamefont {Gauguet},
  \citenamefont {Weatherill}, \citenamefont {Jones},\ and\ \citenamefont
  {Adams}}]{Maxwell2013}%
  \BibitemOpen
  \bibfield  {author} {\bibinfo {author} {\bibfnamefont {D.}~\bibnamefont
  {Maxwell}}, \bibinfo {author} {\bibfnamefont {D.~J.}\ \bibnamefont {Szwer}},
  \bibinfo {author} {\bibfnamefont {D.}~\bibnamefont {Paredes-Barato}},
  \bibinfo {author} {\bibfnamefont {H.}~\bibnamefont {Busche}}, \bibinfo
  {author} {\bibfnamefont {J.~D.}\ \bibnamefont {Pritchard}}, \bibinfo {author}
  {\bibfnamefont {A.}~\bibnamefont {Gauguet}}, \bibinfo {author} {\bibfnamefont
  {K.~J.}\ \bibnamefont {Weatherill}}, \bibinfo {author} {\bibfnamefont
  {M.~P.~A.}\ \bibnamefont {Jones}}, \ and\ \bibinfo {author} {\bibfnamefont
  {C.~S.}\ \bibnamefont {Adams}},\ }\href {\doibase
  10.1103/PhysRevLett.110.103001} {\bibfield  {journal} {\bibinfo  {journal}
  {Phys. Rev. Lett.}\ }\textbf {\bibinfo {volume} {110}},\ \bibinfo {pages}
  {103001} (\bibinfo {year} {2013})}\BibitemShut {NoStop}%
\bibitem [{\citenamefont {Whitlock}\ \emph {et~al.}(2010)\citenamefont
  {Whitlock}, \citenamefont {Ockeloen},\ and\ \citenamefont
  {Spreeuw}}]{Whitlock2010}%
  \BibitemOpen
  \bibfield  {author} {\bibinfo {author} {\bibfnamefont {S.}~\bibnamefont
  {Whitlock}}, \bibinfo {author} {\bibfnamefont {C.~F.}\ \bibnamefont
  {Ockeloen}}, \ and\ \bibinfo {author} {\bibfnamefont {R.~J.~C.}\ \bibnamefont
  {Spreeuw}},\ }\href {\doibase 10.1103/PhysRevLett.104.120402} {\bibfield
  {journal} {\bibinfo  {journal} {Phys. Rev. Lett.}\ }\textbf {\bibinfo
  {volume} {104}},\ \bibinfo {pages} {120402} (\bibinfo {year}
  {2010})}\BibitemShut {NoStop}%
\end{thebibliography}
\end{document}